\documentclass[preprints,article,accept,oneauthor,pdftex,10pt,a4paper]{Definitions/mdpi}

\firstpage{1}
\pubvolume{xx}
\issuenum{1}
\articlenumber{5}
\pubyear{2019}
\copyrightyear{2019}
\history{Received: 30.  10.  2018; Accepted: 09.01.2019; Published: 12. 01. 2019, Symmetry 2019, 11(1), 81}
\graphicspath{{fig/}}
\usepackage{graphicx}
\usepackage{hyperref}
\DeclareGraphicsExtensions{.pdf,.png,.jpg}
\usepackage{amssymb,amsmath,amsfonts}
\Title{The Ambiguity in the Definition and Behavior of the Gravitational and Cosmological `Coupling Constants' in the Theory of Induced Gravity}

\Author{Farkhat Zaripov}

\AuthorNames{Farkhat Zaripov}

\address[1]{N. Lobachevsky Institute of Mathematics and Mechanics,
Kazan Federal University, Kazan, Russia; farhat.zaripov@kpfu.ru  }
\abstract{This work is the extension of author`s research, where the modified theory of induced gravity (MTIG) is proposed. The theory describes two systems (stages): Einstein (ES) and ``restructuring'' (RS). We consider equations with quadratic potential that are symmetric with respect to scale transformations. The solutions of the equations obtained for the case of spaces defined by the Friedman-Robertson-Walker metric, as well as for a centrally symmetric space are investigated. In our model arise effective gravitational and cosmological ``constants'' , which are defined by the ``mean square'' of the scalar fields.
In obtained solutions the values of such parameters as ``Hubble parameter'', gravitational and cosmological ``constants'' in the RS stage fluctuate near monotonically evolving mean values. These parameters are matched with observational data, described as phenomena of dark energy and dark matter.
The MTIG equations for the case of a centrally symmetric gravitational field, in addition to the Schwarzschild-de Sitter solutions, contain solutions that lead to the new physical effects at large distances from the center. The Schwarzschild-Sitter solution becomes unstable and enters the oscillatory regime. For distances greater than a certain critical value, the following effects can appear: deviation from General relativity and Newton's law of gravitational interaction, antigravity.}


\begin{document}
%%%%%%%%%%%%%%%%%%%%%%%%%%%%%%%%%%%%%%%%%%
%% Only for the journal Gels: Please place the Experimental Section after the Conclusions

%%%%%%%%%%%%%%%%%%%%%%%%%%%%%%%%%%%%%%%%%%

\section{Introduction}
This work is related to research in the field of the theory of gravity and cosmology in connection with existing problems given below.

1. The difference in the values of the cosmological constant obtained from astrophysical observations and predictions of the general relativity theory (GRT), taking into account the quantum effects of vacuum polarization, is known in science as the ``problem of the cosmological constant'' (see~\cite{Weinberg C.(1989)}). The acuity of this problem reinforces the fact that this difference is huge  $10^{120}$. 

2. There is a problem of   ``accuracy of measurement of the gravitational constant'' $G$ \cite{Speake(2014),Rosi(2014)}.
For~example, in the International System of Units (SI), for 2008: $G$ = 6.67428 $\times$ 10$^{-11}$ m$^3$ c$^{-2}$ kg$^{-1}$; 
the value of the gravitational constant was obtained in 2000 (Cavendish Experiment): $G$ = 6.67390 $\times$ 10$^{-11}$; in 2010, the value of $G$ was corrected: $G$ =6.67384(80) $\times$ 10$^{-11}$; in 2013 a group of scientists from the International Bureau of Weights and Measures: $G$ = 6.67545$\times$ 10$^{-11}$;  in 2014, the value of the gravitational constant recommended by CODATA became: $G$ = 6.67408 $\times$ 10$^{-11}$; in 2014 the journal Nature published an article by Italian and Dutch physicists, which presented the results of the $G$ measurements, using atomic interferometers: $G$ = 6.67191 $\times$ 10$^{-11}$.

 The recently published results of the new measurements \cite{Luo(2018)} show that, despite two independent methods of measuring the gravitational constant (using torsion pendulum experiments with the time-of-swing method and the angular-acceleration-feedback method), the results differ in the fourth order after the decimal point. The G values of $6.674184 \times 10^{-11}$ and $6.674484 \times 10^{-11}$ were obtained with a relative standard uncertainties of 11.64 ppm and 11.61 ppm, respectively. New measurements make the situation more confusing! In fact, $G$ is not determined even with an accuracy of the fourth decimal place.

Recent observations used (by Dr. Adam Riesz's groups) to calculate the Hubble constant value result in a discrepancy between the results obtained by the Hubble Space Telescope (HST) and Planck observatory \cite{Riess(2018)}.
The Hubble Space Telescope is tuned to measure the parallax Milky Way Cepheid variables and the distances are 1.7--3.6  kpc (the modern Universe).
The measurements of the Planck spacecraft correspond to distant galaxies (the early Universe is about  375,000  years old).
In 2018, the accuracy of the measurement of $H_0$ is increased to $2.3$ percent, which gives $H_0$ = 73.48 $\pm$ 1.66 km $\cdot$ s$^{-1}$ Mpc$^{-1}$.
In the early Universe, based on the data received from the ``Planck'' spacecraft and $\Lambda CDM$ theory, the predicted value is $H_0$ = 67.0 $\pm$ 1.2 km $\cdot$ s$^{-1}$ Mpc$^{-1} $.
The difference is about 9 percent. The accuracy of the measurements is about 4.5 percent.
There is also a variance in the observations made at different times and different methods.
For example, as indicated in the work \cite{Riess(2016)}, the local and direct definition of $H_0$ gives $H_0$ = 73.24 $\pm$ 1.74 km $\cdot$ s$^{-1}$ Mpc$^{-1}$, and the most recent value from  \cite{Planck(2016)}  in consent with  $\Lambda CDM$ is 66.93 $\pm$ 0.62 km $\cdot$ s$^{-1}$ Mpc$^{-1}$.
In our opinion, the problem can be reduced to a strong binding of calculations of the Hubble parameter $H_0$ to the $\Lambda CDM $ model.
In our work we present a model where, due to the oscillatory regime in the solutions of equations, the Hubble parameter also fluctuates with respect to the mean value---which is also a function of time.

3. The problems of so-called ``dark energy'' (DE) and ``dark matter'' (DM). The first of them can be reduced to the problem of existence and smallness of the ``cosmological constant'' (par. 1).  The challenge posed by the cosmological constant problem \cite{Weinberg C.(1989)} has spurred many attempts at directly modifying Einstein's gravity at large distances \cite{Rham(2007)}. As  example of such infrared (IR) modification is the $DGP$ brane-world model \cite{Dvali(2000),Dvali(2001),Dvali(2007)}.  In this scenario, our visible world is confined to a brane in an infinite $5D$ bulk.  As shown in Ref. \cite{Ravanpak(2016), Capozziello(2018)}, the current Planck data used is best suited to the model in a non-planar $\Lambda DGP$.
 In \cite{Arkani(2002),Dvali(2000),Dvali(2001),Dvali(2007),Rham(2007)}, put forward the idea that if gravity is sufficiently weakened in the infrared, then vacuum energy could effectively decouple from gravity or degravitate over time. In our work, we investigate a similar mechanism associated with the nonlocal behavior of a gravitational system with scalar fields in the classical approximation.

An attempt is also made in \cite{Mishra(2017),Sahni(2000),Peebles(2003),Padmanabhan(2003),Sahni(2006),Copeland(2006),Bousso(2008)} to move from the paradigm of a particle-like WIMP dark matter to an alternative possibility that DM could have the structure of a scalar field. In \cite{Kallosh(2013),Kallosh(2015)}, Kallosh and Linde drew attention to a new family of superconformal inflationary potentials, subsequently called $\alpha$-attractors. In the works mentioned above it is common to use scalar fields to describe DE and~DM.

Our theory is a phenomenological model used for comparison with observational data DE and DM.
Within the framework of modified theory of induced gravity (MTIG), proposed in the works~\cite{Zaripov(2007),Zaripov(2014),Zaripov(2017)}, we attempted to solve the above problems based on the idea of the existence of macroscopic parameter of the theory $(X,X)=X^{A}X^{B}\eta_{AB}\equiv Y$, which generates both gravitational and cosmological~``constants'':
\begin{equation} \label{Es4}
k_{eff} = \pm\frac {w c^{3}} {16\pi\xi (X, X)\hbar}\equiv G_{eff}\frac{c^3}{8 \pi \hbar}, \quad \quad \Lambda_{eff}
= \frac{1}{2\xi Y} (-B+U_{eff}),\quad n=4,
\end{equation}
 where   $\hbar$ - Planck's constant, $c$ - speed of light,  $B=B_0(n-2)/2 -w \varepsilon_v $, $\varepsilon_v$---vacuum energy $B_0, \quad w,\quad \xi$---constants of the theory, $U_{eff}=U_{eff}(Y)$---effective potential of the theory.

 We are going to compare the experimental value of the gravitational constant $G_m$ with the effective ``gravitational constant'':
\begin{equation} \label{Eg4}
G_m\equiv\frac{8 \pi k_n \hbar }{c^3}\equiv 6.565362\cdot 10^{-65}cm^2=\pm\frac {w } {2\xi C_{m} },
\end{equation}

where $ C_ {m} $ is the current value of the function $ Y = Y (t_m), $ $ k_n = G $ is the gravitational constant of Newton, the value of which is  6.674286 $\times$ 10$^{-8}$ cm$^3$ c$^{-2}$ g$^{- 1}$; $ t_m $ is a time parameter corresponding to the current value (approximately 13.8 billion years).
Similarly, in accordance with astrophysical observational data, the modern value of the cosmological constant is assumed to be equal to $ \Lambda_ {m} \simeq$ 1.27143 $\cdot$ 10$^{-56}$ /cm$^2$.

 Functions $X^{A}=X^{A}(\sigma^{\mu})$, where $A,B=1,2, \ldots , D,\quad
\mu,\nu=0,1,\ldots,n-1$, represent $n$-dimensional Riemannian manifold $M$
described by the metric $g_{\mu\nu},$ into $D$-dimensional flat space -
time $\Pi$ with the metric $\eta_{AB}$ \cite{Zaripov(2007)}. For further calculations we set $n=4$.

For a cosmological model (a similar model is constructed for a centrally symmetric space as~well), the mechanism proposed by us reduces to the fact that the differential equations describing the evolution of the functions $Y(t)$ and $a(t)$ have the form
$$\dot{Y}\cdot(\Phi_1(Y, \ a))=0;  \quad  \Phi_2(Y, \ a)=0,$$ where $\Phi_1(Y, \ a)\quad  \Phi_2(Y, \ a)$ some expressions of functions  $Y(t)$ and the cosmological scale factor $a(t)$ and their derivatives up to the second order. For $Y(t)=const$, the second equation goes to the equation matching with the equation in general relativity, and the first equation disappears. Thus,  there are solutions that can both match and and not match with the solutions of the standard theory of gravity. Then the fundamental ``constants'' of theory, such as gravitational and cosmological, can evolve in time, and also depend on coordinates.  In a fairly general case, the theory describes two systems (stages): Einsteinian (ES-stage) and ``restructuring'' (RS-stage). This process resembles the phenomenon of a phase transition, where different phases (Einstein's gravitational systems, but with different constants) pass into each other. Perhaps there is no computable description of such transitions. We can only indicate the ``favorable'' points at which such transitions are possible. These are the moments of time when the second derivative of the scale factor $a(t)$ or the first derivative of $Y(t)$ equals zero. In this paper we show that the values of the observed characteristics of the gravitational field are affected not only by the values of the gravitational parameters, but, for the most part, by their derivatives.

In the article \cite{Zaripov(2014)} to solve Problem 1, we considered two mechanisms for reducing the constant part of the vacuum energy $\varepsilon_ {vac}$. In the first variant, the value $\varepsilon_{vac}$ is compensated by other terms $(-B_0+U_{eff})/(2\xi Y) $ from $\Lambda_{eff}$. The reduction of two values imposes requirements on the constants of the theory ($w,  \ \xi, \  C_0, $) to the accuracy of high orders. The second mechanism for reducing the constant part of the vacuum energy reduces to the multiplicative reduction. Its principle is simple and is based on the law of conservation of energy in phase transitions corresponding to different stages of universe evolution and the structure of the theory which  is related by scale invariance. Note that this mechanism for reducing the vacuum energy is analogical to the mechanism for reducing the divergences in the quantum renormalization theory, despite the fact that the theory under consideration is classic. In the work \cite{Zaripov(2014)} the influence of matter in the form of perfect fluid on the behavior of $Y(t)$ was studied. We can say that in the cosmology based on the MTIG, the principle of the ``whole'' Universe (slightly analogous to Mach's principle) is realized to some extent, which reduces to the existence of a certain parameter $Y$, which in turn depends on all material fields and generates physical ``constants''.

\begin{figure}[H]
\centering
\includegraphics[width=7cm]{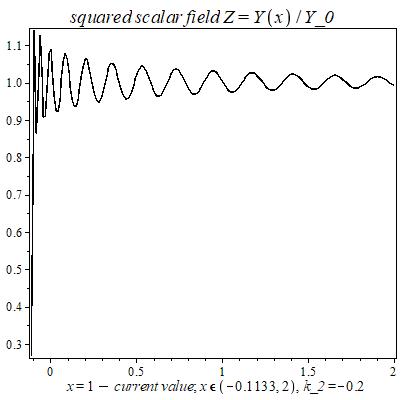}
\includegraphics[width=7cm]{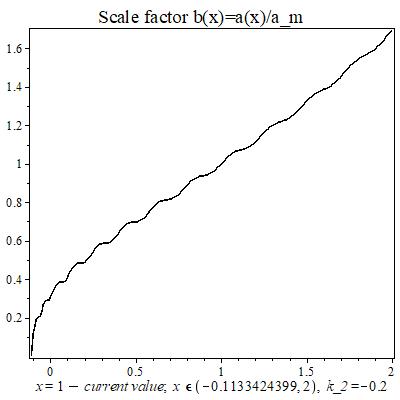}
\caption{{Oscillatory} solutions for flat cosmological model.  $x=t/t_m$, $\tilde{B}_{n}=144.517...$, $k_2=-0.2$. Border conditions: (\ref{nach_us1}), $Z_{m}= 1.0040965.$ } % change ".." in the paper to "...", please confirm if it is right.
\label{kar1}
\end{figure}

\begin{figure}[H]
\centering
\includegraphics[width=6cm]{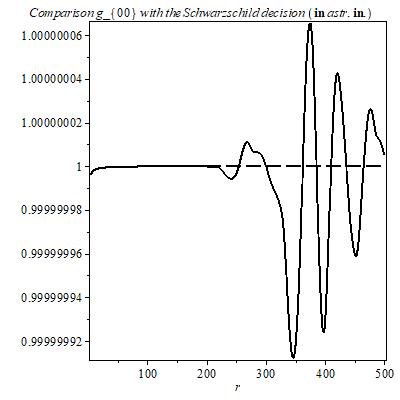}
\caption{The dependence comparison $g_{00}$ (the metric component)  for the centrally symmetric space with the Schwarzschild  de Sitter's  solution (dashed line); r is given in au. $B_{n}=0.0059986...$, $k_1$ = $-$9.263854653 $\cdot$ 10$^{-31}$ au$^{-2}$, $GM = GM_{\bigodot}$. Border conditions: (\ref{sentr13}).}
\label{kar2}
\end{figure}

In this theory we consider the influence of the quadratic, standard potential on the solutions of the RS stage. In our opinion, solutions containing anharmonic oscillations caused by random initial and boundary conditions are of special interest. Unlike the solutions of the Einstein equation with the asymptotics of flat space-time, the presence of a variable ``cosmological term'' leads to a non-local self-interaction of the field $Y$. Fluctuations with a complex spectrum impose on monotonically varying solutions (for example, cited in \cite{Zaripov(2014)}. Such behavior leads to fluctuations in the parameters relative to their mean classical values. Thus, we propose the hypothesis that the value of such parameter as the gravitational ``constant'' $G$, apart from the slow evolution in the RS stage, can fluctuate near the classical value. For example, Figures \ref{kar1} and \ref{kar2} show the normalized numerical solutions $Y(t)/Y_0, \ b\equiv a(t)/a_0, x\equiv t/a_0$, $Z \equiv Y(r)/Y_0$  of equations (cited in this article).

First of all, we require the consistency of the results of the theory (after comparison with observational data) and then look for the predictive possibilities of the theory. Proceeding from this, the choice of the parameters of the theory should lead to fluctuations affecting the fourth (maximum to third) order of the solution $Y(t)/Y_0\simeq 1000m_1m_2\ldots$, interpreted as the modern value of this parameter. Of course, back in time, the values of the parameters could have been substantially different and these parameters implemented qualitatively different stages of evolution (Figure \ref{kar1}). The graph of the scale factor shown in Figure \ref{kar1} can be reconciled with the results of observational data (for example, to calculate the Hubble constant \cite{Riess(2018)}), if the difference of 9 percent obtained by measuring distant (corresponding to the early Universe) and close objects is attributed to averaged local fluctuations of the graph, and 4.5 percent of the inaccuracy \cite{Riess(2018)} can be attributed to the local fluctuations.

Research on numerical solutions for the case of a centrally symmetric space has been made. On the one hand, obtained solutions agree with the observational data, on the other hand, at far distances from the center (more  than 0.01 parsecs from the Sun and more than one kiloparsec from the center of the galaxy) the same solutions lead significantly to another astronomical picture. Even if these examples do not correspond to reality, they demonstrate the existence of models with essentially non-local behavior and with non-flat asymptotics that do not contradict observations.

Solutions comparison of the cosmological model and the model in the case of a centrally symmetric space leads to the fact that DM and DE can be described in a single concept. Please note, this concept does not entirely correspond to the interpretation of ``the evolution of constants''. An important factor is that the energy is hidden in terms containing derivatives of the field $Y$. Due to the phenomenological description, we do not study the quantum mechanical nature of the transformation of DE into DM and vice versa. Such interpretations are possible within the framework of our theory, but all that are the issues for further research.

\subsection{Introduction to the Original Theory}

Historically \cite{Zaripov(2007)}, the theory is based on the generalization of the string theory:

\begin{equation} \label{n1}
S_0 = \frac{1}{w} \int \left\{ - \frac{1}{2}
(\nabla_{\nu}{X},\nabla^{\nu}{X} ) + {\xi} {R} (X,X) + U +L_m (X, S) \right\}
 \sqrt{- g} \hat d^{n} \sigma.
\end{equation}

 In (\ref {n1}) the following notation is used:
$$Y \equiv (X,X)=X^{A}X^{B}\eta_{AB}, \quad
 (\nabla_{\nu}{X},\nabla^{\nu}{X})=\nabla_{\nu}X^{A}\nabla_{\mu}X^{B}g^{\nu\mu}\eta_{AB}, $$ \quad
 where $A,B=1,2, \ldots , D,\quad \mu,\nu=0,1,\ldots,n-1$;  here fixing  the Levi-Civita connection $\nabla$ of the metric $g$;
$U=U(X^{A})$ - is the potential dependent of the fields $X^{A}$.
For simplicity, in this paper $ U (X^{A}) = U (Y (X^{A})) $.  $ L_m (X, S) $ - characterizes all possible interactions $ X^{A} $ with other fields of matter.

In the context of our paper, some of the modified scalar-tensor theories of gravitation can be transformed to the form (\ref{n1}), without taking into account the Einstein term which is absent in (\ref{n1}). The structure of the theory can also be rewritten in the framework of the modern theories $ f (R)$ \cite{De Felice(2010),Nojiri(2007),Nojiri(2014),Peter(2016),Luongo(2018),Aviles(2012),Sakharov(1968),Visser(2002),Andrianov(2006),Linnemann(2018),Scholz(2011),Aalbers(2013),Dengiz(2011),Carballo-Rubio(2015),Kamenshchik(2016),Bars(2016),Grin(1988),Regge(1977),Paston(2010),Sheykin(2014),Stephani(2003),Bamba(2012)}.
In Ref. \cite{Rham(2008),Nojiri(2007),Nojiri(2014)} were presented some classes of modified gravity, considered as the gravitational alternative to dark energy. In Ref. \cite{Luongo(2018),Peter(2016)}, the authors  ``...revised the cosmological standard model presuming that matter, i.e., baryons and cold dark matter, exhibits a nonvanishing pressure mimicking the cosmological constant effects''. The authors use an approach in which dark energy arises as a consequence of the thermodynamics of the Universe \cite{Peter(2016)}.

For the first time, the concept of ``induced gravity'' was introduced in the work of A.A. Sakharov in 1967, in Ref. \cite{Sakharov(1968)}. The basic idea is that gravity is not “fundamental” in the sense of particle physics. Instead it was argued that gravity (general relativity) emerges from quantum field theory (see Refs. \cite{Visser(2002),Linnemann(2018)}).  The one loop effective action automatically contains terms proportional to the cosmological constant, the Einstein–Hilbert action, plus “curvature-squared” terms. At present, the concept of induced gravity has expanded. For example, this concept is used in the theory of superstrings, branes, for the transition to the 4-dimensional theory of gravity.  The~paper~\cite{Andrianov(2006)} presents research of the non-compact (4 + 1)-dimensional fermionic model with a strong local four-fermion interaction and an additional induced background gravitational field. The gravity is generated completely by five-dimensional matter and therefore gravity is induced in the Zel’dovich--Novozhilov--Sakharov~sense.

In the most of these works, including ours, the Weyl interpretation is considered as the geometric component of the physical theory, see the extensive review by  \cite{Scholz(2011)}.
The Weylian metric on a differentiable manifold M (in Ref. \cite{Scholz(2011)} $dim M = 4$) can be given by pairs $(g, \ \phi)$ and of a non-degenerate symmetric differential two form $g$, here of Lorentzian signature (3, 1), and a differential 1- form $\phi$. The Weylian metric consists of the equivalence class of such pairs, with  $(\tilde{g}, \tilde{ \phi })$$\sim$$(g, \phi)$  iff
 \begin{equation} \label{weil}
(a):\tilde{g} = \exp(2 \psi ) \ g; \quad (b):\tilde{\phi}=\exp(W  \psi) \ \phi
\end{equation}
for a real function $\psi$ on $M$.   A change of representative (\ref{weil}) is called Weyl or scale transformation; it consists of a conformal rescaling (a) and a scale gauge transformation (b).  Also, such a conformal rescaling is called a change of frame. A manifold with the Weylian metric $(M, [g, \  \phi])$ will be called the Weylian manifold. A quantity $\phi$ has conformal weight  $W$ if, under the Weyl transformation, it transforms via (b).  Examples are: $W(g_{\mu \nu})= 2,  \ W(g^{\mu \nu}) = -2$ etc.
A theory or an expression invariant under this transformation is called conformally invariant, or is said to possess Weyl invariance or Weyl~symmetry.

 Jordan-Brans-Dicke (JBD) theory assumes a scalar field $\chi$ of scale weight $W(\chi)=-1$, coupled to gravity (a pseudo-Riemannian metric $g$) by a Lagrangian of the following type \cite{Scholz(2011)}
\begin{equation} \label{brans}
L_{JBD}= \left\{R \chi - \frac{w}{\chi} \nabla_{\nu}{\chi} \nabla^{\nu}{\chi} + L_m  \right\}
 \sqrt{- g},
\end{equation}
with a free parameter $w$ and scalar curvature $R$.

The “original” one (defining the affine connection as the Levi-Civita connection of the Riemannian metric) like in (\ref{brans}) is called Jordan frame. The one in which the scalar field (and thus the coefficient of the Einstein-Hilbert term, the gravitational coupling coefficient) is scaled to a constant is called Einstein frame \cite{Scholz(2011)}.

More recent literature (Fujii/Maeda 2003, Faraoni 2004), prefers a slightly different form of the scalar field and the Lagrangian, $\phi^2=\chi \xi^{-1},$  $\xi =1/(8 w)$, $W(\phi)=-1$,
\begin{equation} \label{far}
L_{F}= \left\{\xi R \phi^2 - \frac{1}{2}  \nabla_{\nu}{\phi} \nabla^{\nu}{\phi} + L_m  \right\}
 \sqrt{- g}.
\end{equation}

 Penrose (1965) showed that $L_{F}$ is conformal invariant for $\xi=\xi_0$ ($n$ spacetime dimension).

\begin{equation} \label{ns2}
\xi_0 \equiv -\frac{n-2}{8(n-1)},\quad \xi \equiv \xi_0 -\frac{\delta_{\xi}}{4(n-1)}.
\end{equation}

Generalization of the action (\ref{far}) on $D$ of scalar fields ($\phi \rightarrow X^{A}, \ \phi^2 \rightarrow Y \equiv (X,X) $) and taking into account the Polyakov`s action for bosonic strings (first term in action (\ref{n1}) for $n = 2$)  brings us to action~(\ref{n1}). The reverse transition to the Einstein frame or the frame of the Brans-Dicke  is ambiguous due to the presence of the  term $(\nabla_{\mu}X, \nabla^{\mu} X)$. Considering that conformal invariance in principle does not allow one to distinguish one energy scale from another, we  chose \cite{Zaripov(1995)} conformally invariant generalization of string theory for multidimensional objects. Thus this theory avoids the problem of uncertainty of quantum field theory on the Planck scale. For example,  Friedmann-like cosmological models can be described in this framework.

Conformal invariance is a tempting but problematic component for theories of gravity. The motivations for invoking it are mainly quantum-theoretical: an opportunity for a renormalizable theory, a better understanding of black hole entropy and perhaps even a step further along the road to a theory of everything. Conformal symmetry plays a critical role in string theory. The Einstein-Hilbert action of general relativity is not conformally invariant and would have to be modified.  One of these options is through the action (\ref{n1}) for $\xi = \xi_0$.

Many modifications of scalar-tensor theories are associated with the study of conformal transformations in the theory of gravity \cite{Aalbers(2013)}.
The Jordan-Brans-Dicke theory also carries the basic features of a Weyl geometric structure.
However, conformal symmetry has to be broken at lower energies, to account for the obvious existence of massive particles.  The theory with Weyl symmetry breaking is given in Ref. \cite{Dengiz(2011)}. When the Weyl symmetry is broken, the graviton gets a mass in analogy with the Higgs mechanism (Ref. \cite{Dengiz(2011)}).

Thus, in the “induced gravity” approach, the initial classical action of GR is equated to zero---the idea of the zero Lagrangian.
For example, a similar approach is used in Ref. \cite{Kamenshchik(2016)}, where was proposed new version of the description of the crossing of singularities.  It is based on the transitions between the Jordan and Einstein frames.

In paper \cite{Bars(2016)} some physical problems related to the existence of an antigravity regime are analyzed, and the possibility of the indirect observation of such a phenomenon are discussed. It was emphasized that using Weyl invariance one can get a geodesically complete theory.

The author of paper \cite{Carballo-Rubio(2015)} bases  the discussion on the gravitational theory known as Weyl transverse gravity. ``General relativity exhibits the well-known cosmological constant problem while in Weyl transverse gravity the cosmological constant sector is protected due to gravitational scale invariance, and this is possible as the result of abandoning the assumption of full diffeomorphism invariance''. In this connection, we note that it is precisely the scale invariance of the theory that is the cause of the multiplicative contraction of the component of the gravitational constant, even in the classical version of the theory considered in paper \cite{Zaripov(2014)}.

 Thus through action  (\ref{n1}), within the framework of the Weyl manifold concept, $D$ of scalar fields $X^{A}$ is considered. At the same time, these same fields are considered as the coordinates of some ambient flat pseudo-Euclidean space $\Pi$. This is also known as nonlinear sigma model.
The presence of the Weyl conformal symmetry (for $\xi=\xi_0$) is an important motivation for the use of the action (\ref{n1}).

After varying action (\ref{n1}), the field equations for $g$ have the following form (See Refs. \cite{Zaripov(2007),Zaripov(2014)})):

\begin{equation} \label{Ten}
T_{(tot)\alpha\beta} \equiv T_{\alpha\beta} + T_{(e)\alpha\beta}=0,
\end{equation}
where $T_{\alpha\beta}$ and $T_{(e)\alpha\beta}$---the Energy--Momentum Tensors (EMT) of fields $X^{A}$ and other fields of matter (for example, perfect fluid), respectively. In the bosonic string theory ($n=2, \ \xi=\xi_0, \ T_{(e)\alpha\beta}=0$), Equations (\ref{Ten}) are the called constraint Equations \cite{Grin(1988)}. If $(X,X)=const$, Equations (\ref{Ten}) are similar to the Einstein equations with an effective gravitational constant.

In the general case, to solve Equations (\ref{Ten}) is a very difficult task, even in the case of conformal invariance.
Instead, we can set a narrower task  finding all solutions of Equations (\ref{Ten}) for the case of conformally flat metrics. Then for the conformally invariant case the metric can be reduced to  $(h)$ - the metric of  Minkowski spacetime. By means of transformations (\ref{weil}): $(g, X) \rightarrow (h, \tilde{X})$.
 The equations are~simplified:

 \begin{equation} \label{Ten2}
(\nabla_{\nu} \tilde{X}, \nabla_{\mu} \tilde{X}) - \frac{1}{6} [ \nabla_{\mu} \nabla_{\nu} \tilde{Y} + h_{\mu \nu} (2 U + (\nabla_{\alpha} \tilde{X},\nabla^{\alpha} \tilde{X}))]=0,
 \end{equation}

  \begin{equation} \label{Ten3}
 \Box \tilde{X}^{A} + 4 \Lambda_X \tilde{Y} \tilde{X}^{A} = 0,
 \end{equation}

where all covariant derivatives are taken by metric (h); $n=4$; $U=\Lambda_X \tilde{Y}^2$; $\tilde{Y}=(\tilde{X}, \tilde{X})$.

Considering that conformally flat spaces cover a large enough range, we can try to apply the quantization procedure for Equations (\ref{Ten2}) and (\ref{Ten3}) in the Minkowski spacetime. If there is no self-action ($\Lambda_X=0$) or $Y=const$ the equations for the fields $X^A$ are linear. There are problems associated with the dimension $D$ and boundary conditions..

Let's mark the important fact, that for strings ($n = 2$) the general solution of the constraint equations has the form:
\begin{equation}\label{gtti}
B_0g_{\mu \nu} =  (\nabla_{\mu} X, \nabla_{\nu} X) \quad \mu, \nu =
\overline{0,n-1},
\end{equation}
where $B_0$  is arbitrary function.
Thus the metric $g_{\mu \nu}$ is connected by a conformal transformation with the induced metric $(\nabla_{\mu}X, \nabla_{\nu} X)$ on the surface $X^A = X^A(\sigma^ {\mu})$.  When $B_0 = const \neq 0$ $(B_0$~=~1), the Equations (\ref{gtti}) are the conditions for  immersion $M$ in a multidimensional flat spacetime $\Pi$.
Unfortunately, at $n \ > \ 2$, when considering the Equations (\ref{Ten}), in case of arbitrary dimension $n$ this statement  is not true.

However, when considering narrower tasks in the class of conformally flat metrics, the~relation~(\ref{gtti})
is one of the solutions of the Equations (\ref{Ten}). In \cite{Zaripov(2007)} for a class of metrics with the symmetry of a homogeneous isotropic space, an attempt was made to show that solutions (\ref{gtti}) minimize the total energy of the system corresponding to the Equation (\ref{Ten}). Thus, we can use the Higgs mechanism to obtain solutions (\ref{gtti}). Examples of joint solutions of Equations (\ref{Ten}) and (\ref{gtti})  are given in Section \ref{sec2.1}.

Note that the Equations (\ref{gtti}) fix the conformal gauge. The above reasoning led us to the construction of a phenomenological theory, where relations (\ref{gtti}) were taken as the basis or ansatz.  ``The breaking of the original conformal symmetry happened so that the embedding condition (\ref{gtti}) is fulfilled''.  This is the hypothesis of our phenomenological modified scalar-tensor theory, which has an analogy with the Brans-Dicke theory. This theory is an alternative to the theory of GR. However, its equations for the case $ Y = const = C_m $ are completely reduced to Einstein's equations (GR) with a cosmological constant. Based on this, for small deviations of $Y$ from a constant value $C_m$, the difference from the known results of the GR theory will be small---beyond the limits of experimental accuracy (famous experiments), which is achieved by selecting the parameters of the theory. This is what we show in this article. In the case of a space with spherical symmetry, the difference between MTIG and GR will appear at large distances from the center.

 We consider this model as an intermediate stage before formulating a consistent field theory that 250
takes into account the approaches of  local isometric embedding’s methods.  These methods were
considered in  \cite{Regge(1977),Paston(2010),Sheykin(2014),Stephani(2003)}. From  these researches it follows that each manifold requires a separate study on “embedding” , which makes it difficult to develop a general  theory used for comparison with the observed data.

It is known that any $n$-dimensional Riemannian manifold can be locally isometrically embedded in an $D$-dimensional flat ambient space, with $D = n(n + 1)/2$. If a manifold has any symmetries, the number of the ambient space dimensions can be smaller than  $n(n + 1)/2$. The difference between the dimension of the flat ambient space and the dimension of the original manifold is called the embedding class, $p = D - n$. In particular, we have $p = 1$ for constant curvature spacetimes, while for spherically symmetric spacetimes one obtains $p \leq 2$   \cite{Stephani(2003)}.  When $p > 2$ no systematic classification of manifolds has still been performed. Therefore,   the  phenomenological model in which we can distinguish the so-called macroscopic part associated with the parameter $Y$ was proposed. We are trying to isolate some effective equations for the variables $g$ and $Y$, without terms that depend on individual fields $X^A$.

In the context of our article, the ``induced gravity'' means that in the initial action  the Einstein's term $R/(2 \kappa)$ is not explicitly introduced. The introduction of such term , at first violates the conformal invariance of the theory, secondly leads to the instability of known solutions because of the emergence of the effective gravitational ``constant'':
  $$\frac{1}{\kappa_{eff}}=\frac{1}{\kappa}+2 \xi (X,X)\Rightarrow 0.$$

 When it tends to zero, additional singularities arise.  For example, in the case of one scalar field ($\phi$), instability arises for ($\xi=\xi_0$) \cite{Zaripov(1986)}. Although in the mentioned above works  \cite{Kallosh(2013),Kallosh(2015)}  made an attempt to use instability to realize generic chaotic inflation models. In connection with what has been said, it is necessary to point out the problem connected with the sign choice in (\ref{Es4}) and (\ref{Eg4}). Unlike single scalar field $\phi$ (where the analog of the expression $(X,X)$ is $\phi^2>0$), the sign of $(X,X)$ is undefined.
The sign of $G_m$ must be positive. However, before analyzing the solutions of equations, we can not tell the signs of parameters $\xi, C_m=Y(t_m).$

For a system with matter, the following self-consistent equations were obtained \cite{Zaripov(2010)}:
\begin{equation} \label{E1}
G_{\alpha \beta}=\frac{1}{2\xi Y}[-\frac{n-2}{2}B +U]g_{\alpha \beta}+\frac{1}{Y}[\nabla_{\alpha} \nabla_{\beta}
- g_{\alpha\beta} \Box]Y - \frac{w}{2 \xi Y}  T_{(e)\alpha\beta},
\end{equation}
where $G_{\alpha \beta}$---the Einstein tensor; $T_{(e)\alpha\beta}$---the Energy--Momentum Tensor (EMT)
of matter fields (for~example, perfect fluid).

The consequence of these
equations is the law of conservation of energy, which has the form:

\begin{equation} \label{E2}
  -\frac{n-2}{2} \nabla_{\beta}B + \nabla_{\beta}Y \cdot
 (\xi R+\frac{dU}{dY})- w\nabla_{\alpha}T_{(e)\beta}^{\alpha}=0,
  \end{equation}
and the equation on the field $Y$:
 \begin{equation} \label{E3}
\Box Y = \frac{n-2}{4(n-1)\xi}[-nB + 2 \xi R Y +
\frac{2n}{n-2}U]
-\frac{w}{2\xi(n-1)}T_{(e)\alpha}^{\alpha}
\end{equation}

Equations (\ref{E1})  is an analogue of Einstein`s equations for a macroscopic medium.

While deriving the macroscopic equations, the following assumptions are made:

1.  the induced metric (mapping) $(\nabla_{\mu} X, \nabla_{\nu} X)$ is related to the metric of the manifold $M$ by means of Formula (\ref{gtti}). This model allows us to interpret the development of  Universe as development of $n=4$ dimensional objects embedded in the multidimensional flat spacetime $\Pi$.

  2. the equation for scalar fields acquires an additional  term $S^{A}$ due to interaction with vector fields. Then these equations have the form
\begin{equation} \label{Y1}
 \Box X^{A} + 2 \xi R X^{A} + 2 \frac{dU}{dY} X^{A}  = S^{A}.
\end{equation}

The specific form of this term depends on the interaction model.

 From the mathematical point of view, the solution of the inverse task is assumed. Solving the macroscopic equations (\ref{E1}), we find the metric $g_{\mu \nu}$ and the field $ Y $. Then, solving Equations~(\ref{gtti}) and~(\ref{Y1}) we find  $ X^A, S^A $. Our approach is similar to the method of finding the unknown potentials given in papers \cite{Chervon(1997),Zhuravlev(1998)}.
%\begin{em}

\begin{Remark}
 In the context of suggested consideration of macroscopic equations,  the following considerations can be offered concerning conditions (\ref{gtti}). In  general case if we make the following substitution
$(\nabla_{\mu} X, \nabla_{\nu} X)=B_0 g_{\mu \nu} k_{\mu \nu}  \quad \mu, \nu =
\overline{0,n-1},$
where  $k_{\mu \nu}$ - some tensor functions , then  the resulting equations will have the same form as Equations (\ref{E1})--(\ref{E3})and the EMT matter will be redefined:
\begin{equation} \label{gtti2}
T_{(e)\alpha\beta} \Rightarrow T_{(e)\alpha\beta} +\frac{1}{w}(k_{\alpha \beta} -\frac{1}{2} g_{\alpha\beta} k_{\mu \nu} g^{\mu \nu} ).
\end{equation}
\end{Remark}

Then we can assume that the “embedding” condition (\ref{gtti}),  does not limit the proposed theory, but changes the EMT.  In further research, we intend to consider the problem of deriving covariant Equations (\ref{E1}) directly from action (\ref{n1}).

In order to take into account the effect of vacuum polarization energy into gravity, we highlighted from EMT matter a part related to this energy, which satisfies the equation of state: $\varepsilon_{vac}+p_v=0$, where $\varepsilon_{vac}$ and $p_v$  are interpreted as the energy density and vacuum polarization pressure. Therefore,  in the equations (except for (\ref{gtti}))  we made a substitution:$B_0\Rightarrow B,$
\begin{equation} \label{opr_B}
B=\frac{n-2}{2}B_0-w \varepsilon_{vac}.
\end{equation}

The action (\ref{n1})  has the property of conformal invariance
if  $\xi=\xi_0$, ($\delta_{\xi}=0),$  $ U(X^{A})=U_{0}\equiv \Lambda (X,X)^{2}$, where for dimension $n=4$: $\delta_{\xi}=-12 \xi -1$.
This invariance is expressed in the fact that the equations obtained by
  varying action (\ref{n1}) with respect to the fields $\hat g$ and
  $\hat X$ are invariant under the local Weyl scale changes
\begin{equation} \label{cp}
g_{\mu\nu} \Rightarrow \exp(2\psi) g_{\mu\nu}, \qquad X^{A}
\Rightarrow \exp(4\xi_0 (n-1)\phi)X^{A},
\end{equation}
for an arbitrary function $\psi = \psi(\sigma^{\mu})$.

The condition  (\ref{gtti}) for $B_0 = const$ limits the conformal invariance to scale transformation\linebreak ($\psi = const$). Indeed, if in (\ref{cp}) we substitute
$n=4,\quad \exp(\psi)=b=const$ then $g_{\mu\nu} \Rightarrow b^2 g_{\mu\nu}, \quad Y \Rightarrow Y/b^2,\quad B_0 \Rightarrow B_0/ b^4$ (similarly for $B$).
The equations do not change for an arbitrary value of the coefficient
$\xi$. Thus, fixation of the function $B_0$ (in the general case, this parameter can be a function of the coordinates) leads to scale fixation of the theory. In our notation, $ B_0 $ is a dimensionless quantity. The fields $X^{A}$ interpreted as the coordinates of the space $M,$ have the dimension of a centimeter, which implies $[Y]=cm^{2}$ and $[w]=cm^{4}$. The action  (\ref{n1}) is a dimensionless quantity ($\hbar=1,\quad c=1 $).

To  harmonize general parameters of the cosmological model and the model of galaxies, the dependence of the coupling constants of the theory on the energy scale (temperature) of the Universe is important. Quantum effects are taken into account by the one-loop renormalization group for coupling constants.
For many theories with conformal coupling, the conformal case: $ \xi = \xi_0$ corresponds to a fixed point of the one-loop renormalization group for the constant $\xi$  \cite{Buchbinder(1992)}.

The authors of  \cite{Gorbunov(2014)} investigated the possibility that the additional dark radiation has an origin associated with the scale invariance.

Note that conformal transformations that preserve the condition (\ref{gtti}) are not limited to the case of scale transformations. We can verify that in addition to them, the theory is invariant under transformations: $g_{\mu\nu} \Rightarrow b^2 g_{\mu\nu},$ $Y \Rightarrow Y/b^2,\quad B_0\Rightarrow B_0/ b^4$, when $b=Y b_0, \quad b_0=const$, where $\xi=\xi_0, \quad n=4$.
Proceeding from this fact, the terms with derivatives of $ B $ were left in the Equations~(\ref {E1})--(\ref {Y1}). Here this topic is not touched upon and it is assumed that $ B = const $.

\begin{Remark}

The action (\ref{n1}) and the Equations~(\ref{E1})--(\ref{Y1})) were obtained in the earliest articles \cite{Zaripov(2007),Zaripov(2017)}, on the basis of the signature $(- + + +)$ of the space $\Pi.$
In this article, we carried out some subsequent calculations on the basis of the opposite signature $(+ - - - )$. However, in order not to confuse the reader, the original notation was retained in the old version, although the signs of the potential coefficients, also the sign of $B$, must be changed.
\end{Remark}

In genera case, we get the systems of
``macroscopic'' Equations (\ref{E1})--(\ref{E3}), ``microscopic''
(\ref{Y1}) and constraint Equations (\ref{gtti}). The study of complete system of equations requires the definition of the model, i.e.,
definition of the functions $S^{A}$. The fixed sector of the fields $\{X^1,X^2,...X^k\},\quad k<D$ can play the role of Higgs scalar fields. It is proper to consider the function $Y$ as the averaged field (the vacuum mean in the tree approximation) by analogy with the mechanism of spontaneous symmetry breaking (the Higgs mechanism). So the previous formulas should be understood in the following sense: $Y=<0\mid (X, X)\mid 0>\simeq(<0\mid X\mid 0>, <0\mid X \mid 0>)$,
 \begin{displaymath}
B_0g_{\mu \nu} \simeq (\nabla_{\mu}<0\mid X \mid 0> , \nabla_{\nu} <0\mid X \mid 0>) \quad \mu, \nu =\overline{0,n-1}.
\end{displaymath}

The latter can be interpreted in the sense that ``geometry'' is created by vacuum averages.

We note the works of \cite{Rham(2008),Rham(2014)} Claudia de Rham and her colleagues,
where cosmological models with scalar fields, including branes, were studied, taking into account quantum effects (see also review~\cite{Clifton(2012)}).

%%%%%%%%%%%%%%%%%%%%%%%%%%%%%%%%%%%%%%%%%%
\subsection{Different Types of Solutions}

For the ``embedding'' case $(n=4, B_0=const)$, as follows from the Equations (\ref{E2}), the following cases are possible (Zaripov (2010)):

 (I) $Y=C=const,$  $\nabla_{\beta}T_{(e)\alpha}^{\beta}=0.$

 Note also that $Y=C=const, B=const.$  $\Rightarrow$ $\nabla_{\beta}T_{(e)\alpha}^{\beta}=0.$

In this case, we obtain equations that match the Einstein equations, with the gravitational constant $G_{eff}=const$ and with the cosmological constant $\Lambda_{eff}=const$.

Equations (\ref{Y1}) can be rewritten as:

\begin{equation} \label{CY1}
 \Box X^{A} +( 4 \frac{B-U}{C} +2 \frac{dU}{dY}+\frac{w}{C}T_{(e)\alpha}^{\alpha}) X^{A}  = S^{A}.
\end{equation}

For the cosmological model with the EMT of perfect fluid and the potential $ U=U_{0}\equiv \Lambda (X,X)^{2},$ free fields ($S^{A}=0$) $X^A$
acquire mass $\mu$, when $\mu^2=-4\frac{B}{C}
+\frac{w}{C}(\varepsilon -3P),$ where $\varepsilon, P$ is the density of energy and pressure.

 (II) $Y \neq const,$ \emph{and
a separate conservation law for matter is fulfilled}:
 $\nabla_{\beta}T_{(e)\alpha}^{\beta}=0.$  In this case, from~(\ref{E2}) follows equation

\begin{equation}  \label{sl2}
 \xi R+\frac{dU}{dY}=0.
\end{equation}

Equations (\ref{Y1}) can be rewritten as:

\begin{equation}\label{CY2}
 \Box X^{A} = S^{A}\end{equation}
free fields ($S^{A}=0$) $X^A$
 have zero mass.

There is also a third case:

(III) When $Y \neq const, \quad B=const$ \emph{separate law of
 conservation of matter is not necessarily fulfilled}.  This case is a
 generalization of the previous case. The law of conservation takes the form:

\begin{equation} \label{T72}
  \nabla_{\beta}Y \cdot
 (\xi R+\frac{dU}{dY})=w\nabla_{\beta}T_{(e)\alpha}^{\beta}.
\end{equation}

%%%%%%%%%%%%%%%%%%%%%%%%%%%%%%%%%%%%%%%%%%
\section{Cosmological Solutions}

\subsection{Cosmological Vacuum Solutions. Y = const}\label{sec2.1}
Let's consider the above equations under potential
\begin{equation} \label{pot}
U =\Lambda_X (X,X)^{2}+f_{w}(X,X) \equiv \Lambda_X Y^{2}+f_{w}Y ,
\end{equation}
($n=4;\quad \rho=2; \quad B_0=const,$)for the case of a homogeneous, isotropic
cosmological model. A possible mechanism for the appearance of the term $f_ {w}$ is proposed in the article \cite {Zaripov(2014)}.

The metric form of the manifold $\Pi$, has the form
\begin{equation}  \label{met} d s^2 =- d t^2 + a^2 (t)((d
\chi)^2 + K (\chi)d \Omega^2),
\end{equation}
when $K (\chi) =\{ \sinh^2
{\chi};\sin^2 {\chi};\chi^2\}$---respectively for the models of open,
closed and flat types. $d \Omega^2$---is the metric form of a sphere, with a unit radius, expressed in spherical coordinates.

 $Y=C=const.$
 For the case of vacuum ($S^A=0, \quad T_{(e)\alpha\beta}=\varepsilon_{vac}g_{\alpha\beta} $, $B=B_0-w \varepsilon_{vac}$).
 Equations~(\ref{gtti})--(\ref{E2})can be analytically solved.

The equations for the scale factor have the form:

\begin{equation} \label{mas}
 \dot{a}^2(t)=-k + h_0^2  a^2(t),\quad h_0^2= - \frac{\Lambda C^2-B +f_{w} C}{6 \xi C} = \frac{\Lambda_{eff}}{3},
\end{equation}
$k= -1, \quad1,\quad0$ - for open, closed and flat types of spaces, respectively.

The equations for the fields $X^{A}$ take the following form:
\begin{equation} \label{tah}
 \ddot{X} +3\frac{\dot{a}}{a} \dot{X}+ \frac{(3+l)k}{a^2} X-(4 \frac{B}{C} -2 f_{w} )X^{A}=0,\quad l \in N.
\end{equation}

Here, $l$ eigenvalues for the three-dimensional Laplacian $\bigtriangleup_3 X=-(3+l)k.$

 Particular  solutions ($l=0$) of these equations are found \cite{Zaripov(2007)} that satisfy the conditions of ``immersion''  (\ref{gtti}).
For the closed model, these solutions
have the form:
\begin{equation} \label{masshtz}
 a(t)=\frac{\cosh(th_0 )}{h_0}.
 \end{equation}

\begin{equation} \label{zakritx}
X^{5} = \frac{\sinh(t h_0 )}{h_0}, \quad X^a =a(t)k^a.
\end{equation}
where $k^a $- immersion function of  $3$  dimension sphere
 \begin{displaymath}
 k^1 = \sin {\chi} \sin {\theta} \cos {\phi},\qquad k^2 =
\sin {\chi} \sin {\theta} \sin {\phi},
\end{displaymath}
\begin{equation} \label{ka} k^3 = \sin
{\chi} \cos {\theta}, \qquad k^4 = \cos {\chi}.
\end{equation}

 And for the open type of space:
\begin{equation} \label{masshto}
 a(t)=\frac{\sinh(th_0 )}{h_0},
 \end{equation}
\begin{equation} \label{otkr}
X^{5} = \frac{\cosh(t h_0 )}{h_0}, \quad X^a =a(t)k\tilde{}^a,
\end{equation}
 where $k\tilde{}^a$ follows from $k^a$ by replacing $\sin {\chi}, \cos {\chi}$ with $\sinh {\chi}, \cosh {\chi}.$

 For the closed model manifold $\Pi$ forms ``one-sheeted hyperboloid'' in a five-dimensional subspace of flat space $M$ and described by the equation:
\begin{equation} \label{constz}
(X^1)^2+(X^2)^2+(X^3)^2+(X^4)^2-(X^5)^2=\frac{q}{h_0^2}.
\end{equation}

For the case of an open model surface equation has the form:
\begin{equation} \label{constz2}
(X^1)^2+(X^2)^2+(X^3)^2-(X^4)^2+(X^5)^2=\frac{q}{h_0^2}.
\end{equation}

In the derivation of  (\ref{masshto})--(\ref{constz2}) we assumed that the matrix $\eta_{AB}$ (the metric of the space $M$) dimension $D>5$ is diagonal and this diagonal for the closed and open type of space has the form  $(q, q,q, q, -q, q_{1}, ..,q_{D-5})$, $(q, q,q, -q, q, q_{1}, ..,q_{D-5})$.

For the given solutions, the conditions (\ref{gtti}), (\ref{tah}) and $(X, X)=C$  define the relationship between the constants:
  \begin{equation} \label{svyz1}
  \Lambda_X C = \frac {B(3+\delta_{\xi})}{2C}-  \frac {f_w (5+\delta_{\xi})}{4},
\end{equation}
which follows from the requirement of the Equations (\ref{tah}) for the functions (\ref{otkr}).

 Note that when $ f_w=0$ $\Rightarrow$ $\Lambda_{eff}=3 B/C$  - does not depend on $\xi$; and when $\delta_{\xi}=-3$ $\Rightarrow \Lambda_X=0.$

From the requirement (\ref{gtti}) $\Rightarrow$ $q = B_0$.

The condition $(X, X)=C$ for (\ref{otkr}) leads to the relation:

\begin{equation} \label{svyz2}
  C = \frac {q }{h_0^2}= \frac {3 (B_0)}{\Lambda_{eff}}.
    \end{equation}

From (\ref{mas}), (\ref{svyz1}) and (\ref{svyz2}) follows
\begin{equation} \label{svyz4}
   \Lambda_{eff} = \frac {3 B}{C}- \frac {3f_w}{2}=\frac {3 B_0}{C}.
  \end{equation}

From (\ref{svyz5}) and $B=(B_0-w \varepsilon_{vac})$ we get

\begin{equation} \label{svyz5}
\Lambda_{eff} = \frac {3 B_0}{C}, \quad w \varepsilon_{vac}=-\frac {f_w C}{2}, \quad \Lambda_X C=\frac {B_0(3+\delta_{\xi})}{2C}+\frac {w \varepsilon_{vac}}{C}.
  \end{equation}

Thus, if all embedding conditions of the Friedmann world into multidimensional flat spacetime $M$ are met, the cosmological constant does not depend on the polarization energy of the vacuum for the model constructed by us. The result obtained can be used to investigate the case of perturbations.

The above solutions correspond to the special case $S^{A}=0$---without taking into account the interaction of the fields $X^{A}$ with other fields. Then, the case (\ref{svyz5}) corresponds to the minimum of the potential $V_1=(\Lambda_X Y+f_w -B/Y)/(6\xi)$, included in the Equations (\ref{E1}) and  (\ref{tah}) (in the particular case~(\ref{mas})).

There is a problem of defining the numerical values of the parameters of the theory. The number of essential parameters can be reduced to three. For them we use the following notation:

 \begin{equation} \label{min4}
f_n= \frac{f_w}{6 \xi}, \quad L_n=\frac{\Lambda_X C}{6 \xi}, \quad  B_n=\frac{B}{6 C \xi}.
  \end{equation}

So that
\begin{equation} \label{min5}
L_n+f_n -B_n=-\frac{ \Lambda_{eff}}{3}.
\end{equation}

In order to generalize the theory for $S^{A}\neq 0$ and to reduce the number of parameters of the theory, we will not be limited to the model described by the solutions (\ref{mas})--(\ref{svyz5}). For this we adopt the following arguments.

For minimum potential energy: $\dot{V_1}=0 \Rightarrow  \Lambda_X Y_{min}=-B/Y_{min}$ $\Rightarrow$ $V_{1 min}=-(2 B/Y_{min} - f_w)/(6\xi).$ Taking into account Equation (\ref{mas}), let's assume $V_{1 min}=-\Lambda_{eff}/3$ $\Rightarrow$
\begin{equation} \label{min1}
f_n= 2 B_n-\frac{\Lambda_{eff}}{3}, \quad L_n=-B_n, \quad  C\equiv Y_{min}.
  \end{equation}

Instead of selecting ansatz (\ ref{min1}) due to the small value of the observed cosmological constant $\Lambda_{mod}=\Lambda_{eff}$ - at $t=t_0,$ for $|\Lambda_X| C^2 \ll |B| $ two other ansatzes were considered as well:
\begin{equation} \label{minn4}
f_n= 2 B_n-2 \frac{\Lambda_{eff}}{3}, \quad L_n=-B_n+\frac{\Lambda_{eff}}{3}, \quad  C\equiv Y_{min}-
  \end{equation}
 corresponds to the minimum potential $V_2=(\Lambda_X Y^2+f_w Y -B)/(6\xi)$;
\begin{equation} \label{min2}
f_n= 2 B_n, \quad \Lambda_X =0
  \end{equation}
this case  is interesting because it is possible to obtain analytical solutions of differential equations, even if the substance is present in the form of perfect fluid.

%%%%%%%%%%%%%%%%%%%%%%%%%%%%%%%%%%%%%%%%%%
\subsection{Cosmological Solutions with Matter}
\unskip
In the article \cite{Zaripov(2017)} a phenomenological model was proposed. The model of the interaction of the field $Y$ and matter in the form of perfect fluid, with the density of energy and pressure:
 \begin{equation} \label{Plot}
 \varepsilon=\varepsilon_{r0}/a^4 +\varepsilon_{p0}/a^3+ [(Yf_{r1}+Y^2 f_{r2})/a^4+ (Yf_{p1}+Y^2 f_{p2})/a^3].
 \end {equation}
\begin{equation} \label{Plot2}
 p=\varepsilon_{r0}/(3 a^4) + [(Yf_{r1}+Y^2 f_{r2})/(3 a^4)].
 \end {equation}

Equations (\ref{E1}) and (\ref{E2}) take form :
 \begin{equation} \label{E4}
\lambda^2=-\frac{\dot{Z}}{Z}\lambda
-\frac{k}{b^2}-Z (\tilde{L}_{n} +\tilde{F}_2)- \tilde{f}_{n}  -\tilde{ F}_1 + \frac{\tilde{B}_{n}}{Z}
-\frac{\tilde{E }}{Z},
\end{equation}

 \begin{equation} \label{Ex4}
\dot{Z} \{\dot{\lambda}+2 \lambda^2
+\frac{k}{b^2}+ 2 Z (\tilde{L}_{n}+\tilde{F}_2)+ \tilde{f}_{n} + \tilde{ F}_1  \}=0.
\end{equation}

For convenience in computer modeling, dimensionless variables are introduced
 \begin{equation} \label{par1}
x=\frac{t}{t_m};\quad b=b(x)=\frac{a(x)}{a_m};\quad
Z=Z(x)=\frac{Y(x)}{C_0},\quad \lambda=\lambda(x)=\frac{\dot{b}}{b}
\end{equation}
where $t$---proper time , dot denotes the derivative by  $x$; $C_0=Y(t_1)$ some value of the field $Y$, which we associate with a constant solution $Y=const$, discussed above in particular ($C=C_0$), $a_m$---dimension value $cm^2$---it is convenient to correspond to the modern value of the scale factor or the age of the universe. In the first case  $b(t_m)=1$ corresponds to the modern value of the scale factor, and $t_m$---to the age of the universe. Such a scale is convenient if the required functions are expressed through the scale factor. However, we do not know the modern value of the scale factor, but we assume  $t_m$$\sim$13.7~$\cdot$~10$^{9}$ years. Therefore, when the desired functions are expressed in terms of time, we select $t_m$ for the parameter $a_m$.  Then $x=1, \ b(1)=b_m$ correspond to the modern values of the parameters.

The last equation (taking into account the previous one) for $\dot{Z} \neq 0$ can be rewritten as:
 \begin{equation}  \label{Exx4}
\dot{\lambda}= \frac{k}{b^2}+\frac{2\dot{Z}}{Z}\lambda+\tilde{f}_{n}+\tilde{ F}_1-2\frac{\tilde{B}_{n}}{Z}+2 \frac{\tilde{E }}{Z}.
\end{equation}

Here we have introduced the following notation:
$$\tilde{L}_{n}=L_{n} t_m^2,\quad \tilde{f}_{n}=f_{n} t_m^2,\quad \tilde{B}_{n}=B_{n} t_m^2;$$
$$\tilde{E}=-\frac{\rho_{p}}{b^3} -\frac{\rho_{r}}{b^4};  $$
$$\tilde{ F}_1=-\frac{\mu_{p1}}{b^3} -\frac{\mu_{r1}}{b^4},\quad \tilde{F}_{2}=-\frac{\mu_{p2}}{b^3} -\frac{\mu_{r2}}{b^4},$$
and also re-parameterized constants taking into account their dimensions
$$\rho_{p}=-\varepsilon_{p0} w t_m^2 /(6\xi C_0 a_m^3),\quad \rho_{r}=-\varepsilon_{r0} w t_m^2 /(6\xi C_0 a_m^4);$$
$$\mu_{p1}=-f_{p1} w t_m^2/(6\xi a_m^3),\quad \mu_{r1}=-f_{r1} w t_m^2/(6\xi a_m^4);$$
$$\mu_{p2}= -f_{p2} C_0 w t_m^2/(6\xi a_m^3),\quad \mu_{r2}=-f_{r2} C_0 w t_m^2/(6\xi a_m^4).$$

It is interesting to compare the equations obtained from (\ref{E4}) and
 (\ref{Exx4}), for $Z=const = Z_0$ with
Einstein's equations with the same EMT:

 \begin{equation} \label{EE4}
\lambda^2+\frac{k}{b^2}=\gamma
+\frac{8\pi}{3}G_{eff}|\tilde{\varepsilon}|,
\end{equation}

 \begin{equation} \label{EExx4}
\dot{\lambda}=\frac{k}{b^2}- 4\pi G_{eff} |(\tilde{\varepsilon} + \tilde{p})|,
\end{equation} where

 $$G_{eff} = |\frac {w}{16 \pi\xi C_0 Z_0 }|,  \quad
\gamma= \frac{\tilde{B}_{n}}{Z_0}-\tilde{L}_{n} Z_0-\tilde{f}_{n};  \quad \tilde{\varepsilon}=-\varepsilon / a_m,  \quad \tilde{p}=-p / a_m. $$

The first of these
Equations ((\ref{E4}) and (\ref{EE4})) will match, and Equation (\ref{Ex4}) disappears. Equivalent (at~$\dot{Y} \neq 0$ ) to Equation (\ref{Ex4}), Equation (\ref{Exx4}) does not match to Equation (\ref{EExx4}). We recall that in the case of Einstein's equations the second one is a differential consequence of the first.

Thus, as already indicated in previous works, in the proposed model, the evolution of the universe contains two stages that were named as ``Einstein'' (ES-stage)---when $\dot{Y}=0$ and ``restructuring'' (RS-stage)  when $\dot{Y}\neq 0$. This process resembles the phenomenon of a phase transition, where different phases (Einstein's gravitational systems, but with different constants) pass into each other.

From a mathematical point of view, at any time the solutions of the Equations (\ref{E4}) and~(\ref{Ex4})---describing the ES and RS stages, can pass into each other. To describe such solutions it is necessary to join functions of the scale factor $b(t)$  and the field $Z(t)$ and their first derivatives at the point $t_1$---corresponding to the moment of transition.
These transitions are similar to the first-order phase transitions and apparently can be used to describe transition from the inflationary phase to the next phase \cite{Zaripov(2014)}:
\begin{equation} \label{uslov1}
 a_e(t_1)=a_r(t_1); \ \dot{a}_e(t_1)=\dot{a}_r(t_1); \ Y(t_1)=Y_0; \ \dot{Y}(t_1)=0,
 \end{equation}
where the index ``e'' denotes the solutions $\dot{Y}(t)=0$ or the corresponding ES-stages, and the index ``r''---RS-stages.

  Transitions similar to the second-order phase transitions are described by the system, if~the conditions (\ref{uslov1}) are supplemented by the condition of equality of second derivatives at the transition~point:
 \begin{equation} \label{uslov2}
 a_e(t_1)=a_r(t_1); \ \dot{a}_e(t_1)=\dot{a}_r(t_1); \ \ddot{a}_e(t_1)=\ddot{a}_r(t_1); \ Y(t_1)=Y_0; \ \dot{Y}(t_1)=0; \ \ddot{Y}(t_1)=0.
 \end{equation}

A necessary condition for the existence of solution $Z(t)=Z_0= const$ (if $Y_0=C_0$ then \scalebox{0.95}[0.95]{$Z_0=1$})---equations describing both ES and RS stages, is the fulfillment of the following conditions on the model~parameters:
\begin{equation} \label{uslov3}
Z_0 \tilde{f}_{n}=2 \tilde{B}_{n};  \ \mu_{p1} Z_0-\rho_p +3 \mu_{p2} Z_0^2 =0; \ \mu_{r1}+ 2\mu_{r2}Z_0=0.
\end{equation}
%%%%%%%%%%%%%%%%%%%%%%%%%%%%%%%%%%%%%%%%%%
\subsection{The Case without Quadratic Terms}

In the articles \cite {Zaripov(2014),Zaripov(2017)} in order to obtain analytical solutions, we consider a linear approximation of $Y$, so that
\begin{equation} \label{usl1}
 \tilde{L}_{n}+\tilde{F}_2=0.
 \end{equation}

 In this paper we want to focus on the existence of nonstandard solutions related to the branching effect of solving equations.

The Equation (\ref{Ex4}) integrated and reduced to the form:
\begin{equation} \label{En2}
\dot{Z}\left(\frac{\dot{b}^2}{b^2}+ \frac{k}{b^2}+\frac{\tilde{f}_{n}}{2}-\frac{2\mu_{p1}}{b^3}- \frac{F_0+\mu_{r1}  ln((b/b_0)^2)}{b^4}\right)=0,
\end{equation}
where
 $F_0, \ b_0$---integration constants.

We can prove that for the case of $\dot{Z}\neq 0$, the scale factor is the solution of Equation
 (\ref{Ex4}), taking into account (\ref{usl1}), and where $Z=Z(t)$ is found by the formula:
  \begin{equation} \label{EH4}
  Z=\dot{b}\left(c_2+\int\frac{b}{\dot{b}^3}[\tilde{B}_{n} -\tilde{E}]db \ \right),
   \end{equation}
  that follows from (\ref{E4}) and (\ref{Ex4}).

  Let`s consider so-called ``equilibrium state'' in more detail. This state is obtained by applying the conditions (\ref{uslov3}) and the additional condition on the constant $F_0$: $F_0\cdot Z_0=\rho_{r}$.  From (\ref{uslov3}), taking into account (\ref{usl1}), follows $\mu_{r1}=0$. These conditions are obtained from the requirement of existence and matching solutions $Z=const$ of Equations (\ref{E4})  and (\ref{En2}).

  After substituting these values of the parameters, besides the condition $\mu_{r1}=0$, the equations (for $\dot{Z}\neq 0$)can be reduced to the form

 \begin{equation} \label{urraf1}
\dot{Z}\left(\frac{\dot{b}^2}{b^2}+ \frac{k}{b^2}+\frac{1}{Z_0}\left(\tilde{B}_{n}-\frac{2\rho_{p}}{ b^3}- \frac{\rho_{r}}{ b^4}\right)-\mu_{r1} \frac{ ln((b/b_0)^2)}{b^4}\right)=0,
\end{equation}
	
 \begin{equation} \label{urraf2}
 \dot{Z}\left(\frac{\dot{Z} \dot{b}}{b}+ \frac{Z-Z_0}{Z_0} \left( \tilde{B}_{n}+\frac{\rho_{p}}{b^3}+ \frac{\rho_{r}}{ b^4}\right)+Z \mu_{r1} \frac{  ln((b/b_0)^2)-1}{b^4}\right)=0.
 \end{equation}

It is of interest to study the influence of the logarithmic term in the Equations (\ref{urraf1}) and (\ref{urraf2}) on their solutions, so we left this term---as some perturbation violating the solutions of the ``equilibrium~state''.

The solution of the Equations (\ref{urraf1}) and (\ref{urraf2}), in the case $\mu_{r1}=0$, is found by the formula:
  \begin{equation} \label{urraf3}
  Z(x)= c_2\cdot \dot{b} + Z_0,  \ c_2=const.
\end{equation}

The function $b=b(x)$, is defined as the solution of the Equation (\ref{urraf1}) (for $c_2\neq 0$). The Equation~(\ref{urraf2}), taking into account  (\ref{urraf3}) and (\ref{urraf1}), becomes their differential consequence.

Surprisingly, the solution for the scale factor does not depend on the constant $c_2$. This equation has the form:
 \begin{equation} \label{urraf4}
\frac{\dot{b}^2}{b^2}=- \frac{k}{b^2}+\gamma_0+\frac{2\rho_{p}}{Z_0 b^3}+ \frac{\rho_{r}}{Z_0 b^4},
\end{equation}
where $\gamma_0= -\frac{\tilde{B}_{n}}{Z_0}$ - defines the cosmological constant (from which, presumably, follows $\tilde{B}_{n}/ Z_0< 0$). As a consequence, it follows from the Equations (\ref{E1}), (\ref{urraf4}) and the solution (\ref{urraf3}) that there are two ``gravitational constants'':  $\mid\frac {w } {2\xi Y_{0} } \mid$ - cosmological gravitational constant and  $\mid\frac {w } {2\xi Y(t) } \mid$---time-dependent function, (possibly) contributing to the gravitational interaction between massive bodies. In addition, the solution (\ref {urraf3}) is noteworthy by the fact that the transition between ES and RS stages takes place at the point when the first derivative $Z(t)$ and the second derivative of the scale factor $\dot{Z }(t_1)=c_2 \ddot{b}(t_1)=0$ equal zero. Thus, the transitions between the stages will be located in the vicinity of the special points ($\ddot{b}(t_1)=0$) for the scale factor function. From this point of view, it is interesting to study all special points, including the equilibrium points $\dot{b}(t_1)=0, \ \ddot{b}(t_1)=0$. In this regard, I want to refer to the work \cite{Lerner(2018)} where he gives arguments to the consistency of the static universe, from the point of view of observational data on galaxies. In the next section, we present a model of a quasistatic universe, where the scale factor fluctuates with respect to a constant value.

At  $Z=Z_{cr}=const$, Equations (\ref{urraf1}) and (\ref{urraf2}) vanish, and Equation (\ref{E4}) takes form:

	\begin{equation} \label{urraf5}
\frac{\dot{b}^2}{b^2}+ \frac{k}{b^2}+ \tilde{B}_{n}\left(\frac{2}{Z_0}-\frac{1}{Z_{cr}}\right) - \frac{\rho_{p}}{b^3}\left(\frac{1}{Z_0}+\frac{1}{Z_{cr}}\right)-\frac{\rho_{r}}{Z_{cr}b^4}-\frac{\mu_{r1}}{b^4}=0.
\end{equation}

 Joining of solutions (\ref{urraf5}) with (\ref{urraf1}) and (\ref{urraf2}) at some point $t=t_{cr}$ is performed at the equality of functions $b(x)$, $Z(x)$  and their first derivatives. As for the continuity of the first derivative of a function, this function is not directly related to the four-dimensional geometry, although in section III  we were able to interpret this parameter as the "radius" of the four-dimensional hyperboloid (embedded in the five-dimensional space-time). In the vicinity of the transition point, we require the continuity of the function $b(x)$, as well as its first derivative. This requirement is associated with the requirement of energy conservation. We can separately consider the question of the continuity of the second derivatives of these functions. Then we can prove the following relations at the transition point $x=x_{cr}$:
\begin{equation} \label{urraf6}
\frac{\dot{Z}_{r}\cdot \dot{b}_{r}}{Z \cdot b}=\frac{\dot{b}^{2}_{e}-\dot{b}^{2}_{r}}{b^2}; \ \frac{\ddot{Z}_{r}}{Z}=2 \frac{\ddot{b}_{e}-\ddot{b}_{r}}{b}-\frac{\dot{b}^{2}_{e}-\dot{b}^{2}_{r}}{b^2}.
\end{equation}

For greater clarity, Equation (\ref{urraf2}) is reduced to the form:
\begin{equation} \label{urraf7}
\dot{Z} \cdot\dot{b}=(Z-Z_0) \cdot \ddot{b}- \frac{Z_0\cdot \mu_{r1}}{b^3} (ln(b^2/b^2_0)-1)
\end{equation}

Let's denote by $F(x)$ the solution (\ref{urraf2}).

Let's consider the following conditions for joining solutions at the transition point $x=x_{cr}$:
\begin{equation} \label{urraf8}
Z(x_{cr})\equiv F(x_{cr})=Z_{cr}; \  \dot{Z} \cdot \dot{b}_{r} \equiv (Z_{cr}-Z_0)\cdot \ddot{b}_{r}-\frac{Z_0 \cdot \mu_{r1}}{b^3_{cr}} (ln(b^2_{cr}/b^2_0)-1)=0,
\end{equation}
where $ b(x_{cr})=b_{cr};$

 In spite of the fact that the values of parameters including the value $\mu_{r1}=0$ were called the ``equilibrium state'', it is interesting to consider a more general case $\mu_{r1}\neq 0$. Moreover, there is a free parameter $b_0$, and we can demand that the term in (\ref{urraf7}) associated with the parameter $\mu_{r1}$ at the critical point $x=x_{cr}$ to be equal zero. To do this, we must select $b_0=b_{cr}/\sqrt{e}$.

\newpage

If we require continuity of $Z(x)$ and its first derivative at $x_{cr}$, it follows from (\ref{urraf8})  that at least one of the equalities holds at the critical point: $Z_{cr}=Z_0$  or $\ddot{b}_{r}(x_{cr})=0$. The solutions $Z_{cr} \neq Z_0$ are of interest. Such solutions define a ``mechanism'' that transfers the constant $Z_0$ (corresponding to the original (ES)) through the intermediate (RS) stage to the new (ES) stage, with the constant $Z_{cr}$. For example, in the case of $\mu_{r1}=0$,  $F(x_{cr})=c_2 \cdot \dot{b}_{r}+Z_0$.  In this case, the relations (\ref{urraf8}) are satisfied even if $Z_{cr}-Z_0=c_2 \cdot \dot{b}_{r} \neq 0$ (when $\ddot{b}_{r}=0$).Thus, transitions from the (RS) stage into (ES) stage are possible in the following form. For points $x < x_{cr}$:  $Z(x)=F(x)$, where $b(x)$ is defined as the solution of the Equation (\ref{urraf1}); at the point $x = x_{cr}$ : $\ddot{b}_{r}=0$  and  $Z_{cr}=Z(x_{cr})$; further for points $x > x_{cr}$ : $Z(x)=Z_{cr}=const$ , where $b(x)$  is defined as the solution of the Equation  (\ref{urraf5}). If $Z_{cr}-Z_0 \neq 0$, then an inverse transition from stage (ES) into RS is possible, for example, at the point where the second derivative of the scale factor (varying by (\ref{urraf5})) $\ddot{b}_{e}=0$.

\begin{figure}[H]
\centering
 \includegraphics[width=0.65\textwidth]{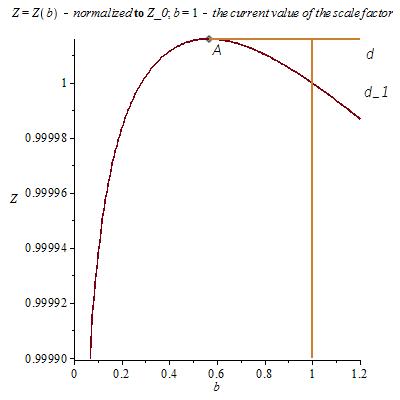}
\caption{Plot of $Z=Y/Y_0$ and $b(t)=a(t)/a_0$:  $b=1 $---current age of the universe. After reaching the point A, the function Y (t) branches. It may evolve on the straight line $d$ or the curve $d_1$. }
\label{kar5}
\end{figure}

\begin{figure}[H]
\centering
\includegraphics[width=0.65\textwidth]{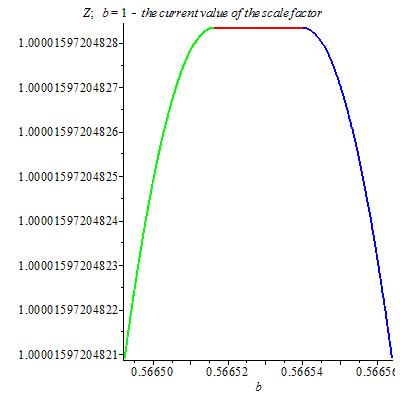}
\caption{Plot of $Z=Y/Y_0$ and $b(t)=a(t)/a_0$:  $b=1 $---current age of the universe. Double transition: from RS into ES and back from ES into RS.}
\label{kar6}
\end{figure}

As an example, Figures \ref{kar5} and \ref{kar6} show the joining of solutions for the function $Z=Z(b)$ describing the transition from RS into ES (Figure  \ref{kar5}) and with the possibility of double transition (Figure  \ref{kar6}). The first point ($A$) is defined by the  moment of time $x_{cr}\simeq 0.512625142$, $b(x_{cr})\simeq 0.5665163348$, where $\ddot{b}_{r}(x_{cr})=0$, $Z_{cr}=Z(x_{cr})$; from the point ($A$) the system can evolve along two trajectories: $d$ - transition into the ES stage or continue to move along the same curve $d_1$ (remain in RS). When the system transits and evolves according to the Equation (\ref{urraf5}), under certain initial conditions, the transition back into RS stage is possible, as shown in Figure  \ref{kar6}. The second transition in Figure  \ref{kar6}, we associate with the point $x_2$, which is defined by the condition $\ddot{b}_{e}(x_2)=0$. The lifetime of the ES stage: $\delta_x=x_2-x_{cr}=0.0000238$.  If we assume that the scale factor is similar to the lifetime of the universe, then $\delta_x$ - is about $400,000$ years. Assuming that the RS stage is currently continuing, the estimate of the rate change of $Z$ (for the model considered in Figure  \ref{kar6}) is $-2.4\cdot 10^{-15}$ per year, and the ``gravitational constant'' increases with the same speed.

The above graphs correspond to the following parameters: $k=0$, $Z_0=1.0001$, $\tilde{B}_{n}/Z_0=-0.7333$, $\mu_{r1}= 0$, $2\rho_p/Z_0= 0.2666$, $\rho_r=0$; and the following initial conditions: $b(0)=10^{-20}$, $ Z=1$ for $b=1$.

However, if the function $Z(x)$  is treated as a macroscopic parameter and only the function $Z(x)$ (not including it`s derivatives) requires the continuity, then transitions are possible at the time when $\dot{b}_{r}(x_{cr})=0$ and the second derivative of scale factor is not necessarily zero. From (\ref{urraf8}) follows, that in the latter case $Z_{cr}=Z_0$ .

As a result, we have possible transitions from RS stage into ES stage of the following form. For~points $x < x_{cr}$:  $Z(x)=F(x)$, $b(x)$ is defined as the solution of the Equation (\ref{urraf1}); at the point $x = x_{cr}$ : $\ddot{b}_{r}=0$ or $\dot{b}_{r}=\dot{b}_{e}=0$  and  $Z_{cr}=Z_0$;  for points  $x > x_{cr}$ :$Z(x)=Z_0=const$ , $b(x)$  is defined as the solution of the Equation (\ref{urraf5}). Similarly, a solution describing the transition from ES into RS can~be~found.

 For the solutions analysis of obtained equations we used qualitative research methods for special solutions of differential equations.  From this analysis follows that in the case of the transition from ES stage into RS and back is more likely near the special points, where the first or second derivative (or~both) of the scale factor equals zero or infinity.

There are numerous versions of modified theories of gravity and cosmology (for example, see the review \cite{Clifton(2012)}). Let us single out some of the works connected with phase transitions in cosmology.
It is possible for a classical field theory to have two stable homogeneous ground states, only one of which is an absolute energy minimum. In the quantum version of the theory, the ground state of higher energy is a false vacuum, rendered unstable by barrier penetration. There is a well-established semiclassical theory \cite{Coleman(1977),Coleman(1980)} of the decay of such false vacuums. In articles \cite{Lee(1987),Hackworth(2004),Masoumi(2016)}, this theory was extended to the inclusion of gravity effects. It is shown that they are not always negligible and may be of critical importance in the late stages of the decay process.

A class of oscillating bounce solutions to the Euclidean field equations for gravity coupled to a scalar field theory with two, possibly degenerate, vacua was studied in articles \cite{Hackworth(2004),Masoumi(2016)}.  In these solutions, the scalar field intersects the vertex of the potential barrier k > 1 times.
The results of our studies do not contradict, but rather agree with the indicated works. However, we paid more attention to gravitational and cosmological effects, which can be compared with observational data related to the phenomenon of dark energy and dark matter, and also we tried to take into account the influence of matter in the form of perfect fluid.
A comparison of the theory considered in articles \cite{Coleman(1977),Coleman(1980),Lee(1987),Hackworth(2004),Masoumi(2016)} with the solutions of our equations (for the case with quadratic potential) shows that our solutions describe the transition from the vacuum $Z_{in}=-\infty $ to the vacuum $Z_0=1$ corresponding to the (for the most part unstable) extremum  of the potential energy.
%%%%%%%%%%%%%%%%%%%%%%%%%%%%%%%%%%%%%%%%%%
\subsection{Oscillating Solutions. Influence of Quadratic Terms}

Let`s consider Equations (\ref{E4}) and (\ref{Ex4}) under the conditions (\ref{uslov3}).
Recall that these conditions are the requirement for existence of common (for both stages) static solutions $Z(x)=Z_0=const$. In doing so, we must understand that the fulfillment of these conditions implements a ``strongly'' nonlinear model.  In this article, we do not claim to develop the final realistic cosmological model, but want to identify the effects associated with nonlinear terms. From the conditions (\ref{uslov3}) let`s express $\mu_{p2}, \ \mu_{r2}$ in terms of $\rho_p, \ \mu_{p1}, \ \mu_{r1}, \ Z_0$  and substitute them in the equations under study that can be reduced to the~form:

\begin{displaymath}
{\frac {d \it b}{d \it t}} \equiv \dot{b} =p,
\end{displaymath}
\begin{equation}\label{furrav1}
{\frac {\dot{p}}{b}}= -2\,Z {\it
L_1}-{\it f_s}   - \frac {p^2}{b^2}
 -\frac {\it k}{b^2}+
 \frac {1}{b^3}\left( {\it m_{p1}}+\,{\frac{2 Z \rho_{p}}{3{{\it Z_0}}^{2}}}-\,{\frac {2{ m_{p1}}\,Z  }{3{\it
Z_0}}} \right)  + \frac {m_{r1}}{b^4} \left( 1-{\frac {\it Z}{ Z_0}} \right),
\end{equation}
\begin{displaymath}
\frac {\dot{Z}}{Z}{\frac {p}{b}} =  - \frac {\it k}{b^2}-\frac {p^2}{b^2}+\frac {1}{b^3}\left( \left( -\frac {Z}{3\it Z_0}+1
 \right) {\it m_{p1}} + \left( {\frac { Z  }{3{\it Z_0}^{2}}}+\frac {1}{Z} \right)  \rho_{p} \right)+
 \end{displaymath}
\begin{equation}\label{furrav2}
 +\frac {1}{b^4}\left( \left( 1  -\frac { Z }{2 \it Z_0} \right){\it m_{r1}}+{\frac { \rho_{r}}{Z }}\right)- {\it L_1}\, Z-{\it f_s}+\frac {\it B_1}{Z}
 \end{equation}

To simplify and reduce the number of parameters, consider the following additional relations that arise from the requirement of the existence of solutions $a=const$ at $Z=Z_0$:
\begin{equation}\label{furrav3}
m_{p1}=-\frac {2}{Z_0} \rho_{p}; \ m_{r1}=-\frac {2}{Z_0} \rho_{r}.
\end{equation}

Also consider two different types of relations, corresponding to the ansatzes examined earlier (\ref{min1}) and~(\ref{minn4}):

\begin{equation}\label{furrav4}
f_s=2\,\frac {B_1}{\it Z_0}+{\it k_2}; \  L_1=-\frac {B_1}{ Z_0^2}.
\end{equation}
\begin{equation}\label{furrav5}
f_s=2\,\frac {B_1}{\it Z_0}+2 {\it k_2}; \  L_1=-\frac {B_1}{ Z_0^2}-{\frac {{\it k_2}}{{\it Z_0}}}.
\end{equation}

Then, the Equations (\ref{furrav1}) and (\ref{furrav2}) are reduced to the form:

\begin{equation}\label{furrav6}
\dot{Z} \left( {\frac {\dot{p}}{b}}+ \frac {p^2}{b^2}+ \frac {\it k}{b^2}-\frac { 2}{ Z_0}\, \left( \frac {\it Z}{ Z_0}-1\right)\left(B_1+ \frac {\rho_{p}}{ b^3} + \frac {\rho_{r}}{ b^4} \right)+K_{rv}\right)=0;
\end{equation}

\begin{equation}\label{furrav7}
\frac {\dot{Z}}{Z}{\frac {p}{b}}+ \frac {p^2}{b^2}  + \frac {\it k}{b^2}-\frac {1}{ Z} \left(\frac { Z}{ Z_0} -1 \right)^2  \left( B_1+ \frac {\rho_{p}}{ b^3} + \frac {\rho_{r}}{ b^4}\right) + K_{ev}=0.
\end{equation}
where $K_{rv}$ and $K_{ev}$ for the case (\ref{furrav4}) equals:$K_{rv}=K_{ev}=k_2$; and for the case (\ref{furrav5}):
 \begin{equation}\label{furrav8}
 K_{rv}=2 k_2 {\it} (1-\frac{Z}{Z_0}), \  K_{ev}=k_2  (2-\frac{Z}{Z_0}).
 \end{equation}

 The parameter $k_2$ has the value of the cosmological constant at the point $Z(t_0)=Z_0$ with the minus~sign.

 Consideration of the ``equilibrium state'' (\ref{uslov3}) led to the appearance of a common factor $ (1-Z / Z_0) $ in the equations. Such an approach resembles the assumption of thermal equilibrium between the effective fluid and the “bath” used in determining the DM mass given in \cite{Luongo(2018)}.

  Let`s consider some numerical solutions of the obtained Equations (\ref{furrav6}) and  (\ref{furrav7}), for the case of a flat space $k=0$ and ansatz (\ref{furrav8}). To define the unknown parameters, we use the following reasoning.
   If in the modern era $t=t_m$ the value of the field $Z(t_m)=Z_m$, then the fractions consistent with observations (similarly to $\Lambda CDM$) of the effective ``cosmological constant'', dust-like and ultra-relativistic matter, as follows from the Equation (\ref{furrav6}), for $RS$ stage equals:

\begin{equation}\label{furrav9}
\Omega_{\Lambda_R}=\frac {2 B_1}{Z_0 H_m^2(1+q_m)}  \left(  {\frac {\it Z_m}{ Z_0}}-1 -K_{rv} \frac {\it Z_0}{ 2 B_1}\right) \simeq 0.739;
\end{equation}

\begin{equation}\label{furrav9a}
\Omega_{p_R}=\frac {2 \rho_{p}}{b_m^3 Z_0 H_m^2(1+q_m)}  \left(  {\frac {\it Z_m}{ Z_0}}-1 \right)\simeq 0.26;
\end{equation}
\begin{equation}\label{furrav9b}
\Omega_{r_R}=\frac {2 \rho_{r}}{b_m^4 Z_0 H_m^2(1+q_m)}  \left(  {\frac {\it Z_m}{ Z_0}}-1 \right)\simeq 0.001.
\end{equation}

By means of $H_m=p/b$ and $q_m$ we denoted the normalized Hubble constant  $H_m=\dot{b}_x/b=t_m \ (\dot{a}_t/a)\simeq 1$ and the deceleration parameter (with the minus sign) $q_m=(\ddot{b}/b)/ H_m^2 \simeq 0.6$, calculated to the moment  $t=t_m$ ($x=1$). Without a loss of generality we can choose $Z_0=1$. Setting $B_1, \ k_2, \ b_m$, from the relations (\ref{furrav9})--(\ref{furrav9b}) we can express $Z_m, \  \rho_{p}, \rho_{r}$  and substitute them in Equations (\ref{furrav6}) and~(\ref{furrav7}).
The relations (\ref{furrav9})--(\ref{furrav9b}) are obtained by analogy with the $\Lambda CDM$ model, but our model differs from $\Lambda CDM$, by the time dependence of the function simulating the $\Lambda$ term  and the presence in the equations of the derivatives of $Z$. If $\dot{Z}\neq 0$, as follows from the Equation (\ref{furrav7}), there is a fraction of the energy associated with the term $\dot{Z}p/(Z b)$, which can have a larger contribution than a term interpreted as a $\Lambda$ term. Also we want to note that in this section we do not investigate the joining of solutions issues of equations describing ES and RS stages, we will study only the solutions of Equations (\ref{furrav6}) and~(\ref{furrav7}) without the multiplier $\dot{Z}$. Although from the results of the previous section it follows that random transitions between the stages are possible at the vicinity of $\dot{Z}=0 $.

Here are some numerical solutions, with the initial conditions:

\begin{equation}\label{nach_us1}
b(1)=1, \ Z(1)=Z_m, \  p(1)\equiv \dot{b}=1 .
\end{equation}

The equations under study are invariant with respect to time shifts $x\rightarrow x+const$. That means that the variable $x$ can also take negative values $x\ni (-\infty, \ \infty)$. For the initial point we will take a point with a singular solution (if it exists).
 As it turned out, the parameters $\tilde{B}_{n}$ and $k_2$ strongly influence the solutions. Taking into account quadratic terms, at $\tilde{B}_{n} > 0$, leads to solutions with anharmonic oscillations, where the ``mean'' oscillation frequency depends on the value $\tilde{B}_{n}$.

\textbf{The Purusha Universe: $k_2=0$}.  Starting from the description sequence, first we consider the exotic case $k_2=0$. A slightly comic name for this model is due to the properties of the solutions, about which, in brief, will be discussed below. The peculiarity of this case (in our opinion is not contrary to the observational data) associated with the fact that the Equations (\ref{furrav6}) and (\ref{furrav7}) for $Z = Z_0$ allow any constant solutions $b(x) =const$.  The initial conditions $\gamma_0: Z_m=Z(x_0),\ b_m=b(x_0), \  P_{b m} \equiv \dot{ b}_m=\dot{b}(x_0)$ that violate this initial state ($Z = \forall Z_0$,  $b(x) = \forall const$), initiate solutions $\gamma: Z=Z(x,Z_m),\ b=b(x,b_m)$  which are related to the initial conditions as some perturbations. The resulting solutions are of stochastic (random) character.The stochasticity is as follows: if from the resulting solution $\gamma$ we choose different point $\gamma_1: Z_n=Z(x_1),\ b_n=b(x_1), \ p_{Z n} =\dot{Z}(x_1), \  p_{b n} =\dot{b}(x_1)$ (corresponding to the moment of time $x_1$) as new initial conditions for the same equations, then the new solutions $\gamma_1$ won`t match $\gamma$. This assertion follows from our computer studies and means that the uniqueness condition for the Cauchy problem is violated.

\begin{figure}[H]
\centering
\includegraphics[width=0.6\textwidth]{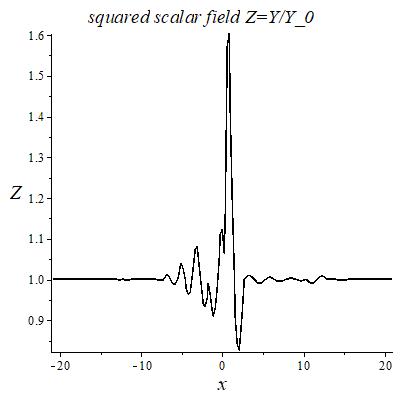}
\caption{Relation of $Z=Z(x)$, where $Z(1)=Z_m= 1.40908675530186$:  $x=t/t_m $---current age of the Universe, $\tilde{B}_{n}=1.44517022939$, $k_2=0$.}
\label{kar7}
\end{figure}

\begin{figure}[H]
 \centering
\includegraphics[width=0.6\textwidth]{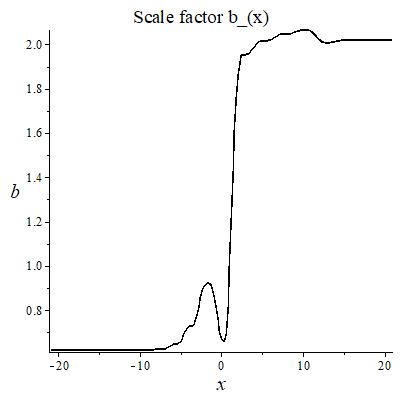}
\caption{Relation of $b=b(x)=a(x)/a_m$,  $b(1)=1 $---the current value of the scale factor. $\tilde{B}_{n}=1.44517022939$, $k_2=0$.}
\label{kar8}
\end{figure}

\begin{figure}[H]
\centering
\includegraphics[width=0.65\textwidth]{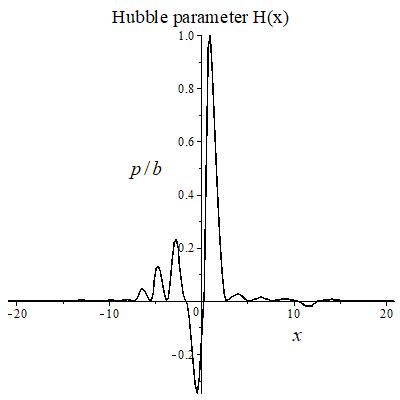}
\caption{Relation of $H(x)=\dot{b}(x)/b(x)$:  $H_m(1)=1$---the current value of the Hubble constant. $\tilde{B}_{n}=1.44517022939$, $k_2=0$.}
\label{kar9}
\end{figure}

\begin{figure}[H]
\centering
\includegraphics[width=0.65\textwidth]{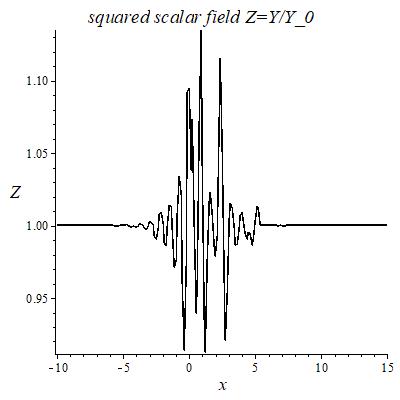}
\caption{Relation of $Z=Y/Y_0$, $Z(1)=Z_m=1.040908675530$:  $x=t/t_m $---current age of the Universe. $\tilde{B}_{n}=14.4517022939$, $k_2=0$.}
\label{kar10}
\end{figure}

\begin{figure}[H]
 \centering
\includegraphics[width=0.65\textwidth]{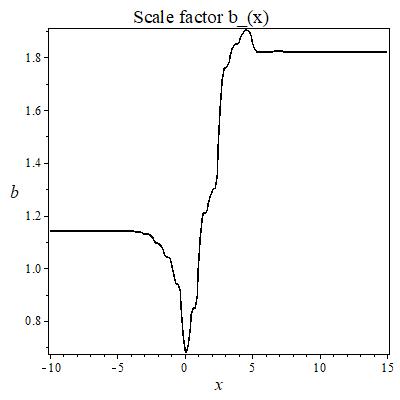}
\caption{Relation of $b(x)=a(x)/a_m$:  $b(1)=1 $---the current value of the scale factor. $\tilde{B}_{n}=14.4517022939$, $k_2=0$.}
\label{kar11}
\end{figure}

\begin{figure}[H]
 \centering
\includegraphics[width=0.65\textwidth]{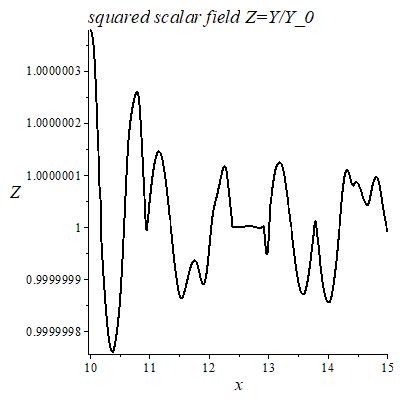}
\caption{Relation of $Z=Y/Y_0$, $Z(1)=Z_m= 1.04072460$:  $x=t/t_m $---current age of the Universe. $\tilde{B}_{n}=14.4517022939$, $k_2=0$.}
\label{kar12}
\end{figure}

\begin{figure}[H]
 \centering
\includegraphics[width=0.65\textwidth]{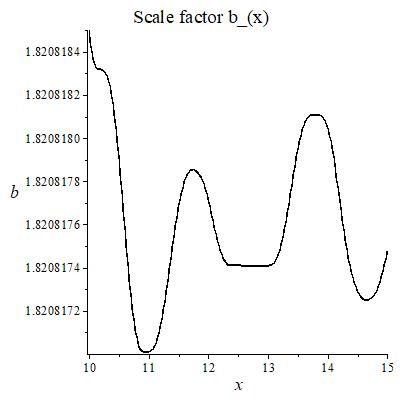}
\caption{Relation of $b(x)=a(x)/a_m$:  $b(1)=1 $---the current value of the scale factor. $\tilde{B}_{n}=14.4517022939$, $k_2=0$.}
\label{kar13}
\end{figure}

\begin{figure}[H]
 \centering
\includegraphics[width=0.65\textwidth]{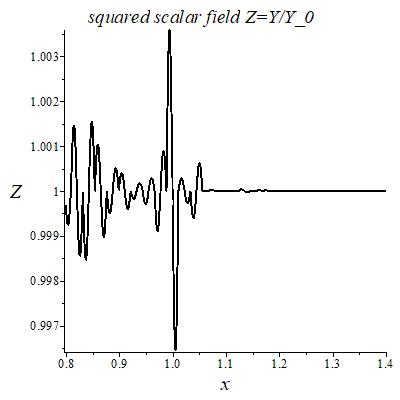}
\caption{Relation of $Z=Y/Y_0$, $Z(1)=Z_m= 1.0000407246$:  $x=t/t_m $---current age of the Universe. $\tilde{B}_{n}=14517.022939$, $k_2=0$.}
\label{kar14}
\end{figure}

To characterize possible solutions, let`s see graphs of numerical solutions, starting with the case $k_2=0$, with the values of the parameters indicated in Figures \ref{kar7}--\ref{kar13}. An increase in the value of $\tilde{B}_{n}$ leads to an increase in the number of fluctuations and a decrease in their amplitudes near the point $x=1$. For comparison, graphs (Figures \ref{kar10} and \ref{kar11}), which differ from (Figures \ref{kar7} and \ref{kar8}) only by a 10-fold increase in the value of $\tilde{B}_{n}$; and Figure \ref{kar14} differs by 10,000 fold increase of $\tilde{B}_{n}$.

In the above solutions (Figures \ref{kar7}, \ref{kar8}, \ref{kar10}, \ref{kar11} and \ref{kar14}), the evolution continues with the transition of the scale factor $b(x)$ (and the function $Z(x)$) to a ``quasi-constant'' value, which is not defined in advance, but rather of random character (depends on the choice of the initial data). In fact,  the parts of graphs seeming to be parts of straight lines,  are not such on small scales. For example, the above graphs Figures \ref{kar10} and  \ref{kar11} when $x\in (10, 15)$ look like Figures \ref{kar12} and \ref{kar13}, that is, the variables change after the seventh decimal.

Interesting features of these solutions are the absence of singularity and infinite extension, as in the model $\Lambda CDM$. The Purusha Universe model has neither the beginning nor the end. Consider, for example, the solution corresponding to Figure \ref{kar8}. For large values of time (for example, for $t>2.3 t_m$) the modern accelerated expansion gradually transforms into fluctuating solution in the vicinity of $a\simeq 2\cdot a_m$. In the past, the scale factor fluctuated around the value $a\simeq 0.62 \cdot a_m$. The amplitudes of the oscillations are small  from $10^{-3}$ to $10^{-4}$ and vary with time. With an increase of $\tilde{B}_{n}$, the amplitude of oscillations decreases, but their averaged frequency increases (other figures). There is a problem of obtaining numerical solutions for large values of $\tilde{B}_{n}$, related to the computer capabilities.

\begin{figure}[H]
 \centering
\includegraphics[width=0.7\textwidth]{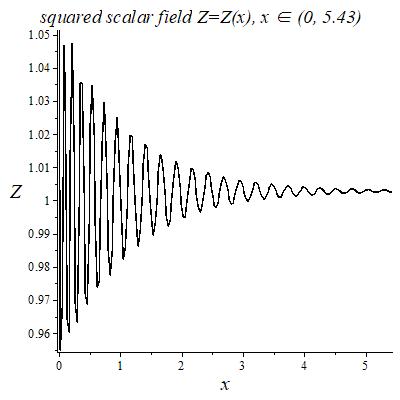}
\caption{Relation of $Z=Y/Y_0$, $Z(1)=Z_m= 1.0041022218...$:  $x=t/t_m $---current age of the Universe. $\tilde{B}_{n}=144.517022939$, $k_2=-0.4$.}
\label{kar15}
\end{figure}

\begin{figure}[H]
  \centering
\includegraphics[width=0.6\textwidth]{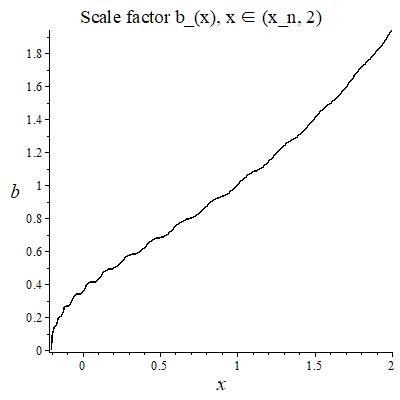}
\caption{Relation of $b(x)=a(x)/a_m$, $Z(1)=Z_m= 1.0041022218...$:  $x=t/t_m $---current age of the Universe. $\tilde{B}_{n}=144.517022939$, $k_2=-0.4$.}
\label{kar16}
\end{figure}

\begin{figure}[H]
  \centering
\includegraphics[width=0.6\textwidth]{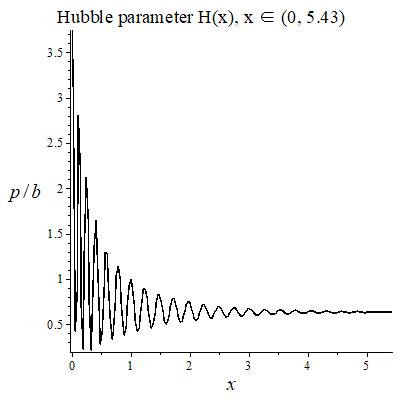}
\caption{Relation of $H(x)=\dot{b}(x)/b(x)$:  $H_m(1)=1$---the current value of the Hubble constant. $\tilde{B}_{n}=144.517022939$, $k_2=-0.4$.}
\label{kar17}
\end{figure}

\textbf{Consideration of the case $k_2 \neq 0$} leads to the models similar to $\Lambda CDM$  in the general scenario of evolution, which are additionally accompanied by fluctuations.
Figures \ref{kar15}--\ref{kar17} show graphs of numerical solutions of equations with initial conditions  (\ref{nach_us1}), for the case with parameters:
\scalebox{0.95}[0.95]{$k_0=0, \  k_1=-0.4$,} $\ B_1=144.517022939$,   $Z_m = 1.0041022218...$,  $\rho_{p}=50.7042299920...$,  \scalebox{0.95}[0.95]{$\rho_{r}=0.1950162692...$} (dots after numbers mean that we give only the first ten numbers after the decimal point).
The main feature of these solutions is the presence of fluctuations that lead to alternation of accelerated with decelerated expansion passing  into a contraction state (at certain values of the~parameters).

For the given model, there is an initial state at certain moment of time $x_{in} \simeq -0.203793929714828..$. The scale factor $b(x)\rightarrow b_{in} \sim 0$ for $x \rightarrow x_{in}$.  It is remarkable that the model has the characteristic properties of the inflationary scenario \cite{Linde(2003)}, perhaps without the scenario of transition from initial $ES$ stage to $RS$ stage proposed in the author's work \cite{Zaripov(2014)}. As follows from computer calculations, for $x\rightarrow x_{in}$ the Hubble function tends to a huge value $H(x)\rightarrow H_{in}>0$ and $Z(x)\rightarrow Z_{in}<0$. For example, at $b(x)= 2.93904...\cdot (10)^{-10}\Rightarrow$ $Z(x)= -4.03515...\cdot (10)^{9}$, $H(x)= 5.62489...\cdot (10)^{23}$. We have not been able to fully explore the model for $b(x)\rightarrow b_{in}$$\sim$0 yet.  This topic of further research. In addition, we must remember that the existence of ``quasi-static'' periods of evolution in our model initiates a new perspective on solving the problem  of large-scale homogeneity and isotropy of the universe. It can be assumed that during the ``quasi-static'' period, the Universe manages to pass into equilibrium state.
%%%%%%%%%%%%%%%%%%%%%%%%%%%%%%%%%%%%%%%%%%

\section{Centrally Symmetric Solutions}

\subsection{Analysis of Equations in Centrally Symmetric Space}

Consider some solutions of the Equations  (\ref{E1}) and (\ref{E2}) for the case of a centrally symmetric space-time defined by the metric:
\begin{equation}\label{sentr1}
ds^2= e^{\nu}dt^2- e^{\lambda}dr^2 -r^2(d\theta^2+sin^2(\theta) d\varphi^2),
\end{equation}
where $\nu=\nu(r), \lambda=\lambda(r)$
For the case of $Y=Y_0= const$---the metric matches with the Schwarzschild-de Sitter metric:

\begin{equation}\label{sentr2}
 \nu_0(r)=-\lambda_0(r); \   e^{-\lambda_0 (r)}=1-\frac{\Lambda_{eff} }{3} r^2-\frac{2GM}{r},
\end{equation}
 where $\Lambda_{eff}$ in the $ES$ stage is expressed in terms of $Y_0$, based on the cosmological solutions discussed in the previous sections:

 \begin{equation}\label{sentr2a}
\frac{\Lambda_{eff}}{3}=- \frac{\Lambda Y_0^2-B +f_{w} Y_0}{6 \xi Y_0}=-L_n Z_1-f_n +\frac{B_n}{Z_1},
\end{equation}
where $Z_1=1$. However, taking into account the possible transitions between the different solutions $Z(r)=const$ considered above, we assume that $Z_1$ is an arbitrary constant solution of $Z(r)=Z_1=const$. For convenience, let`s introduce the constant $K_1$ dependent on the constant $Z_1$:
 \begin{equation}\label{sentr2aL2}
K_1 \equiv K_1 (Z_1)=-L_n Z_1-f_n +\frac{B_n}{Z_1}.
\end{equation}

As is well known, the mass term $2GM$ of the central gravitational field, in the Einstein-Schwarzschild model, arises as a motion integral and is normalized based on comparison with Newton's theory, as well as dimensional consideration $GM = M G_m /c^2$, $G_m$---the current accepted value of the effective gravitational constant $G_{eff}\sim 1/Y=1/(Z_1 C_m)\Rightarrow G_{eff}=G_m/Z_1$ (\ref{Eg4}). Proceeding from our hypothesis about the variable ``gravitational constant'', we assume that the mass of $M$ is constant, and that the value of $G_m$, as a motion integral, changes to  $G_m/Z_1$  ($Z_1=const$).

 In this formulation of the problem we neglect the influence of matter on solutions. To shorten expressions, we will not specify the arguments of functions, and derivative with respect to $r$ will be denoted by ('). % please confirm if it is right.
Unknown functions: $\nu(r), \ \lambda(r), \ Z(r)$.

After the introduction of notation $Z=Z(r)= Y(r)/Y_0$, the Equations (\ref{E1})  in the metric (\ref{sentr1}) take the form:

 \begin{equation}\label{sentr3}
 3 Z  L_n +3 f_n-\frac {3 B_n}{Z } -\frac {e^{-\lambda}}{2 Z }  Z'  \nu'-
 \frac {e^{-\lambda}}{r} \nu' -  \frac {2 e^{-\lambda}}{Z r } Z' - \frac { e^{-\lambda}}{r^2 } +\frac {1}{r^2}  =0
\end{equation}
corresponding to component $G_{1}^{1}$.
 \begin{equation}\label{sentr4}
3 Z  L_n +3 f_n-\frac {3 B_n}{Z }+e^{-lam}  \left(- \frac {{ \nu'}^2}{4} +  \frac { \nu' \lambda' }{4} - \frac { \nu''}{2} -\frac { \nu' Z' }{2 Z}+ \frac { \lambda' Z' }{2 Z}  -  \frac { \nu' }{2 r} +    \frac {\lambda' }{2 r} -\frac {Z'' }{Z} -\frac {Z' }{Z r}  \right)=0
\end{equation}
corresponding to components $G_{2}^{2}=G_{3}^{3}$.
 \begin{equation}\label{sentr5}
3 Z  L_n +3 f_n-\frac {3 B_n}{Z }+e^{-\lambda}  \left( \frac { \lambda' Z' }{2 Z} + \frac {\lambda' }{r} -  \frac {Z'' }{Z} - \frac {2 Z' }{Z r} - \frac { 1}{r^2}   \right) + \frac { 1}{r^2}=0
\end{equation}
corresponding to component $G_{0}^{0}$.

Equation (\ref{E2}), which is the differential consequence of the Equations  (\ref{sentr3})--(\ref{sentr5}) has the form:

\begin{equation}\label{sentr6}
Z' \left( 6Z  L_n +3 f_n+e^{-\lambda}  \left( - \frac {{ \nu'}^2}{4} +  \frac { \nu' \lambda' }{4} - \frac { \nu''}{2} -  \frac { \nu' }{ r} +\frac {\lambda' }{ r}         - \frac { 1}{r^2}   \right) + \frac { 1}{r^2} \right)=0.
\end{equation}

To solve the equations, instead of $\nu(r)$ we use the function $F(r)= \lambda(r)+\nu(r)$. Then the difference of Equations (\ref{sentr3}) and (\ref{sentr5}) is simplified:
\begin{equation}\label{sentr7}
F' \left( \frac {Z' }{2} + \frac { Z}{r} \right) -Z''=0.
\end{equation}

For the case $Z'=0 $ solutions have the form (\ref{sentr2}). For the case $Z'\neq 0$, by simplifying equations by means of algebraic actions, we can obtain three independent equations for the functions $F(r), \  \lambda(r), \ Z(r)$. Two of these are the first-order equations, one is the second-order equation:
\begin{displaymath}
F'= \frac { 1}{Z'r+2 Z} \left[ 2 Z' r \left[\left(  \left( -3 ZL_n-2 f_n + \frac {B_n }{Z}\right) r - \frac {1}{r}\right)e^{\lambda} + \frac {Z'}{Z} -\frac {1}{r} \right]\right]+
\end{displaymath}
\begin{equation} \label{sentr8}
+ \frac { 4 Z}{Z'r+2 Z}  \left(f_n -\frac {2 B_n }{Z} \right) r e^{\lambda} ;
\end{equation}

\begin{displaymath}
\lambda'=\frac { 1}{Z'r+2 Z} \left[ 2 Z' r \left[\left(  \left( -3 Z L_n-2 f_n + \frac {B_n }{Z}\right) r -\frac {1}{r}\right)e^{\lambda} + \frac {Z'  }{Z} +\frac {1}{r} \right]\right]+
\end{displaymath}
\begin{equation} \label{sentr9}
 +\frac { 2 }{Z'r+2 Z}\left[ e^{\lambda}\left( r \left(-3 Z^2 L_n-Z f_n -B_n  \right)- \frac { Z }{r} \right) +\frac { Z }{r}\right];
\end{equation}

\begin{equation} \label{sentr10}
Z''= Z'\left[ \frac { Z'}{Z}+ e^{\lambda} \left[ r \left(-3 Z L_n-2f_n +\frac { B_n}{Z}  \right)- \frac { 1 }{r} \right]- \frac { 1 }{r} \right] +2 Z e^{\lambda} \left(f_n-\frac { 2 B_n }{Z}\right) .
\end{equation}

Recall that the Equations (\ref{sentr3})--(\ref{sentr5}) contain the solution $Z(r)=Z_1= const$, which matches with the solution (\ref{sentr2}) for $Z_1=1$
 \begin{equation}\label{sentr2b}
 \nu_0(r)=-\lambda_0(r); \   e^{-\lambda_0 (r)}=1-K_1(Z_1) r^2-\frac{2GM}{Z_1 r}.
\end{equation}

The Equations (\ref{sentr8})--(\ref{sentr10}) obtained under the condition $Z' \neq 0$ have a constant solution $Z(r)\mapsto Z_2= const$ and this solution differs from the previous one that in the expression (\ref{sentr2b}) $K_1(Z)$ is replaced by another function  $K_2(Z)$:
 \begin{equation}\label{sentr2aL}
e^{-\lambda_0 (r)}=1-K_2 r^2-\frac{2GM}{Z_2 r}, \  K_2 \equiv K_2 (Z_2)=-L_n Z_2-\frac{f_n}{3} -\frac{B_n}{3 Z_2}.
\end{equation}

Thus, in addition to the fact that the equations have multiplier $Z'$, the Equations  (\ref{sentr8})--(\ref{sentr10})  contain two branches of solutions with possible transitions between them. The existence of solution $Z(r)= Z_2= const$, where $Z_2$ is not known in advance but is defined by the previous evolution. This leads to stochastic solutions.

The functions $K_1$ and $K_2$ match, only if the relation $f_n Z_i- 2 B_n =0$. Recall that when considering ansatzes (\ref{furrav4}) and  (\ref{furrav5}), for $Z_0=1$ the value $f_n - 2 B_n $ is $V_{min}=k_1$ and $2k_1$ respectively for ((\ref{min1})~and~(\ref{min2})). We want to obtain the solutions of equations for centrally symmetric gravity confirmed by the observational data. Therefore, let`s  compare the value $k_1$  with the value of observed cosmological constant. Namely, for the reasons given in Section 3 $k_1=-\Lambda_{eff}/3 \Rightarrow K_{1}=-k_1, \ K_{2}=-k_1/3$. Then this parameter has a very small value and as computer calculations show, practically does not affect the solutions near the central mass (when the mass value is large enough). However, it affects solutions when the mass tends to zero.

For further numerical solution of obtained equations, we make the substitution:

\begin{equation} \label{sentr110}
e^{\lambda(r)}=\frac {e^{\alpha(r)}}{f(r)}, \ f(r)=1- K_i r^2-\frac{2GM}{Z_i r}.
\end{equation}

Thus, let us separately allocate the function $f(r)$, which contains singularities at the points $f(r)=0$, associated  with the well-known Schwarzschild-de Sitter solutions.
For the solution  $Z(r)=const\Rightarrow$ $F(r)=0, \ \alpha(r) =0.$

For comparison with observational data, we will estimate the acceleration and velocity of test bodies in a centrally symmetric gravitational field, based on the geodesic equations, which for the metric (\ref{sentr1}) (for the ecliptic plane  $\theta =\pi/2$) can be reduced to the following form:
\begin{displaymath}
\frac {d \varphi}{ds} =\frac {L_{\varphi}}{r^2}, \  L_{\varphi}=const,
\end{displaymath}
\begin{equation} \label{sentr11}
A(r)\equiv\frac {d^2 r}{ds^2} =  \frac {f(r)e^{-\alpha(r)}}{2} \left[  \left( 1+\frac {L_\varphi^2}{r^2} \right) \left(\alpha'(r)-\frac {f'(r)}{f(r)}\right) +\frac {2 L_\varphi^2}{r^3} - e^{(F(r)-\alpha(r))} f(r) F'(r) \left(\frac {d t}{ds}\right)^2  \right].
\end{equation}

From normalization of 4-velocity follows:
\begin{equation} \label{sentr11a}
\left(\frac {d t}{ds}\right)^2 \left(f(r) \left( 1-  \frac {e^{2\alpha(r)-F(r)}}{(f(r))^2}  v_r^2   \right)  \right) = e^{(\alpha(r)-F(r))} \left( 1+\frac {L_\varphi^2}{r^2} \right); \ v_r=\frac {d r}{dt};
\end{equation}

In this paper, because of the complexity of finding solutions we do not solve the equations of geodesics, but  numerically solve the Equations  (\ref{sentr8})--(\ref{sentr10}) and substitute these solutions in (\ref{sentr11}), calculate the acceleration applied by the gravitational field to the stationary test particle with the coordinate $(r, \varphi, 0)$.
After switching to observer time, the right side (\ref{sentr11}) expresses the radial component of the force acting on the test body of the mass=1. In the Newtonian approximation of general relativity, the acceleration of a freely falling body:
 \begin{equation} \label{us}
 A(r)\equiv \frac {d^2 r}{dt^2} = - \frac {\partial\Phi}{\partial r} \approx -\frac {\nu'(r)}{2} e^{\nu(r)-\lambda(r)}.
 \end{equation}

Taking into account that the speed of bodies $v_r\ll c$ (the speed of light), based on the Newtonian approximation, we calculate the ``average'' velocities of so-called "circular" orbits ($v_r\approx0$) (planets,~satellites...):
\begin{equation} \label{sentr12}
v_{co}\simeq \sqrt{-r \frac {d^2 r}{dt^2}} =\sqrt{r \frac {\partial\Phi}{\partial r}}.
\end{equation}

This velocity in Newtonian mechanics is obtained from the equality of the gravitational force and the centrifugal force. In order for the effective gravitational potential  $\Phi$ to characterize only the gravitational field and not to depend of the test body, in the right side of the expression (\ref{sentr11}), in the context of using the Formula (\ref{sentr12}) we need to put $L_{\varphi}=0$.
We present these arguments to justify the calculation $(d t/ds)^2$ from the expression (\ref{sentr11a}) by substituting $v_r=0$.  The smallness of the influence of this term on solutions also follows from computer experiments, for example, calculation of the average orbital velocities of planets in the solar system according to the Formula (\ref{sentr12}) matches with known values with the required accuracy. In the future, the Equations (\ref{sentr8})--(\ref{sentr11a}) should be solved jointly in order to obtain the trajectories of the geodesic.

Thus we look for unknown functions $F(r), \ \alpha(r), Z(r)$. We  consider the boundary conditions as some perturbations to the Schwarzschild-de Sitter solution.  We assume that for $Z=1$ a solution close~to~(\ref{sentr2}) is implemented .Then let the boundary condition reduce to a small difference between $Z$ and its derivative from this solution at the point $r=r_0$.
From a hypothetical consideration, we assume that the point $r_0$ is close to the central mass (the Sun),  but $r_0\neq0$ (to avoid the singularity point at zero).
 For example, we can assume that the perturbations are associated with processes inside a star located at the center of symmetry.  The problem is to determine the value of the $ B_n $ parameter and the values of the boundary conditions.
%-------------------------------------------------------------------------------
\subsection{Numerical Solution of Equations for the ``conditional Sun'' Model}
Based on obtained equations, we will simulate a central gravitational field for a mass equal to the mass of our sun. Proceeding from this, we will calculate all the values in astronomical units ($au$). Then, transferring into this system of units
\begin{equation} \label{sentr120}
k_1 =-9.263854653 \cdot 10^{-31} au^{-2}, \  G M=GM_{\bigodot} \equiv 9.874532 \cdot 10^{-9} au.
\end{equation}

 Thus, the boundary conditions for $r_0=10^{-24} au \approx 1.496 \cdot 10^{-13} m $:
\begin{equation} \label{sentr13}
\alpha(r_0)=0; \  F(r_0)=0;\ Z(r_0)\equiv Z_2=1+243\cdot 10^{-22}; \  Z'(r_0)=10^{-18.7908} au^{-1}.
\end{equation}

After choosing (\ref{furrav4}) from the ansatzes discussed above, we give graphs of numerical solutions for the~parameter
\begin{equation} \label{Bn1}
 B_n=0.005998618729128268 \   au^{-2}.
\end{equation}

The initial values (\ref{sentr13}) and value of $ B_n $ were chosen to some extent randomly, with the requirement that the results of the calculations  do not contradict to the observational data on the one hand and on the other hand lead to new results for distances ($r> 60$ $au$)---for which there are no reliable observational data.

\begin{figure}[H]
 \centering
\includegraphics[width=0.47\textwidth]{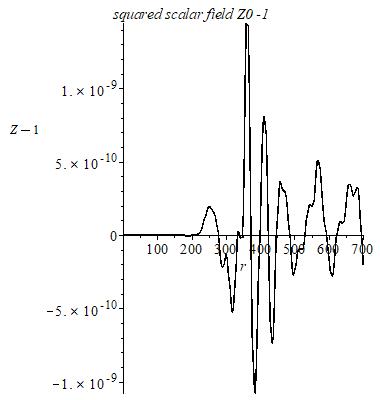}
\caption{Graph of the function  Z(r).
r  is measured in astronomical units. $B_{n}=0.0059986..$, $k_1 =-9.263854653 \cdot 10^{-31} au^{-2}, \  G M=GM_{\bigodot} $.} 
\label{kar18}
\end{figure}

\begin{figure}[H]
  \centering
\includegraphics[width=8cm]{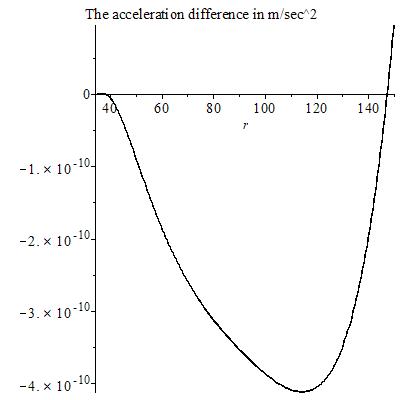}
\caption{Graph of the function ``additional acceleration''  $\bigtriangleup A(r)$.
r  is measured in astronomical units. $B_{n}=0.0059986...$, $k_1 =-9.263854653 \cdot 10^{-31} au^{-2}, \  G M=GM_{\bigodot} $.} % please confirm if the numbers like "1." should be "1" or "1.0", if so, please modify them.
\label{kar19}
\end{figure}

\begin{figure}[H]
 \centering
\includegraphics[width=0.6\textwidth]{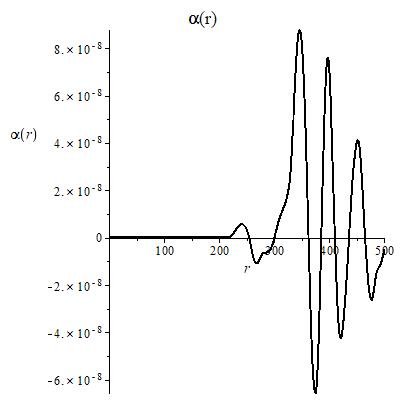}
\caption{Graph of the function $\bigtriangleup \lambda(r)=\alpha(r)$.
r is measured in astronomical units. $B_{n}=0.0059986...$, $k_1 =-9.263854653 \cdot 10^{-31} au^{-2}, \  G M=GM_{\bigodot} $.} % please confirm if the numbers like "2." should be "2" or "2.0", if so, please modify them.
\label{kar20}
\end{figure}

\begin{figure}[H]
  \centering
\includegraphics[width=0.6\textwidth]{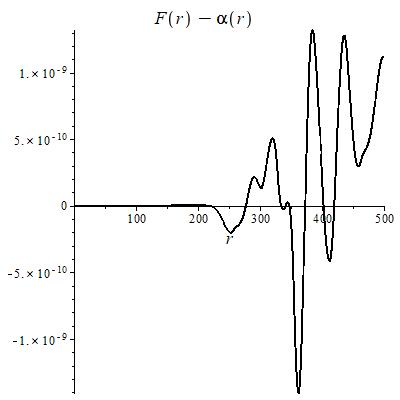}
\caption{Graph of $\bigtriangleup \nu(r)=F(r) - \alpha(r)$.
r  in au. $B_{n}=0.0059986...$, $k_1 =-9.263854653 \cdot 10^{-31} au^{-2}, \  G M=GM_{\bigodot} $.} % please confirm if the numbers like "1." should be "1" or "1.0", if so, please modify them.
\label{kar21}
\end{figure}

\begin{figure}[H]
  \centering
 \includegraphics[width=0.6\textwidth]{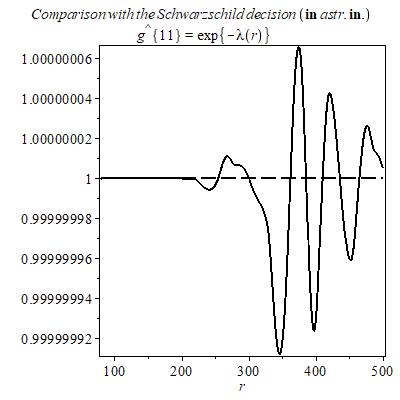}
\caption{A comparison  of the numerical solution $g^{11}$ (solid line) with the Schwarzschild-de Sitter solution (dashed
line); r  in au.}
\label{kar22}
\end{figure}

\begin{figure}[H]
 \centering
\includegraphics[width=0.6\textwidth]{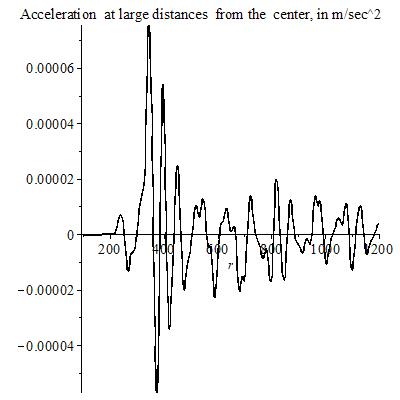}
\caption{This figure shows the acceleration $v_{co}$ vs. distance; r---in astronomical units.}
\label{kar23}
\end{figure}

\begin{figure}[H]
 \centering
\includegraphics[width=0.6\textwidth]{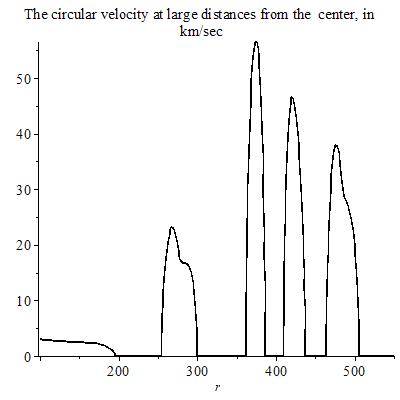}
\caption{This figure shows the ``circular'' orbit velocity $v_{co}$ vs. distance; r---in astronomical units.}
\label{kar24}
\end{figure}

\begin{figure}[H]
 \centering
\includegraphics[width=10cm]{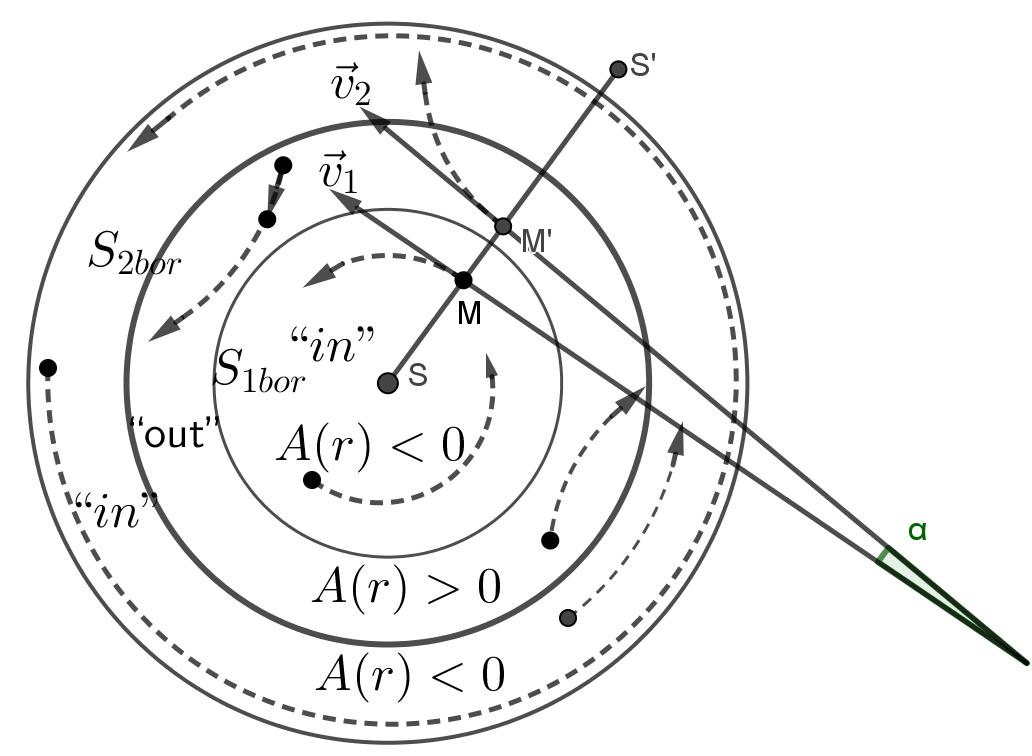}
\caption{The ``gravity'' (``in'') and ``antigravity'' (``out'') bands formation scheme with different directions of acceleration in relation to the center.}
\label{kar25}
\end{figure}

\begin{figure}[H]
 \centering
\includegraphics[width=0.6\textwidth]{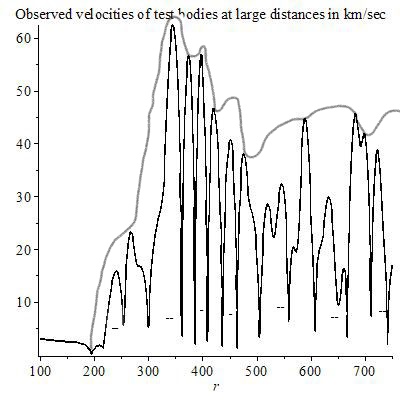}
\caption{``Observed circular rotational velocities'' $v_{co}$ vs. distance circled in pencil, signs {-} (bottom) mean ``antigravity'' ;
 $r \in (100, 700)$ $au$.}
\label{kar26}
\end{figure}

Numerical computations with boundary conditions (\ref{sentr13}) lead to the results shown in Figures~\ref{kar18}--\ref{kar26}.
For distances from the center  $0.01 < r < 60 $ the deviations (difference) of the values  \scalebox{0.95}[0.95]{($g_{00}=e^{\nu(r)}$,  } $g_{11}=e^{-\lambda(r)}$) from the Schwarzschild-de Sitter solutions are $4 \cdot10^{-15}$, $6 \cdot 10^{-14}$, deviations of $Z-1$ are  $10^{-19} \div 6 \cdot  10^{-15}$ (see Figure \ref{kar18}).

For distances from the center $60 < r < 200 $ the maximum deviations of the same variables are  $g_{00}$$\sim$4 $\cdot10^{-12}$, $g_{11}$$\sim$3 $\cdot 10^{-11}$, $Z-1$$\sim$5.4 $\cdot  10^{-12}$.

 Figure \ref{kar19} shows the difference between the accelerations  $\bigtriangleup A(r)=A(r)-A_{0}(r)$  of the test body calculated by the Formula (\ref{sentr11}) for the Schwarzschild-de Sitter metric with the same central mass at $Z=1$. Noteworthy is the appearance of an additional acceleration $\sim$4 $\cdot 10^{-10}$ m/s$^{2}$ to the center (negative sign on the graph indicates direction to the center) for distances $40 < r < 140 $. In this connection, I would like to recall the anomalous acceleration of the ``pioneers'' \cite{Anderson(2002),Nieto(2005)}. The Pioneer anomaly or Pioneer effect was the observed deviation from predicted accelerations of the Pioneer 10 and Pioneer 11 spacecrafts after they passed about 20 astronomical units  on their trajectories out of the Solar System. The apparent anomaly has been a subject of great interest for many years...

 Surprising in our solutions is that the large additional effects (differing from the Schwarzschild-de Sitter solutions) are manifested at large distances from the center  $ r > 200 \ au$. For  $200 < r < 500 $,  the maximum deviations of the same variables are   $g_{00}$$\sim$10$^{-9}$, $g_{11}$$\sim$8 $\cdot 10^{-11}$, $Z-1$$\sim$5$\cdot  10^{-9}$ (Figure \ref{kar18}). The values of the functions  $F(r), \ \alpha(r)$ are approximately $8\cdot  10^{-8}$, and their difference determining the function variation $\nu(r)$ is about $10^{-9}$. The graphs of the functions $\alpha(r), \  F(r) - \alpha(r)$ characterizing the relative changes in the components of the metric $g^{11}, \  g_{00}$  are presented in Figures \ref{kar20} and \ref{kar21}. Figures 2 % please confirm if it is right.
and \ref{kar22} show comparative graphs of the metric components (the Schwarzschild-de Sitter solutions---dashed line).

Thus, at these distances $r> r_{cr}$ (for the considered model of ``solar system'' $ r_{cr}$$\sim$200 $au$) the influence of the terms associated with the central mass decreases and the Schwarzschild solutions become unstable. The solutions become oscillatory, which does not lead to a flat asymptotic  at  $r\rightarrow \infty$. We assume that the asymptotics of the solutions of a centrally symmetric gravitational-scalar field for distances  $r_{cr}\leq r < R_{s} $ can be chaotic.
Here $ R_{s}\sim$10$^{15}$ $au$---corresponds to the de Sitter radius.

In our model, this chaos is defined by the initial conditions, the nature of the occurrence of which is assumed to be random on the one hand. On the other hand, we solve the $RS$ stage equations, which also contains $Z'(r)= 0$ branch similar to the $ ES $ stage (possibly with a small difference between the parameters and solutions). When $r$ is small, the solutions of $RS$ differ very little from $ES$ (the~Schwarzschild-de Sitter solutions), then they represent small random oscillations around $Z(r)=const=Z_1 $ (the influence of $2GM/r$ prevails in $f(r)$). However, for large $ r $, most likely the solutions $ Z (r) = const $ become unstable,  since this branch of the solution is due to a decreasing ``mass~term''~$~2GM/r $.

Let us consider some consequences of our hypothesis. There is the effect of anti-gravity at large distances $r> r_{cr}$.  Figure \ref{kar23} shows the graph of $A(r)$ for $r \in (100, 1200)$, where acceleration regularly changes sign. Due to the fact that  $\bigtriangleup A(r)\sim$$A(r), \ A_{0}(r)\mapsto 0 $, the acceleration of the test body, calculated from the formula (\ref{sentr11}), fluctuates around zero. The acceleration values at some intervals reach up to 7.5 $\cdot 10^{-5}$ m/s$^{2}$. Recall that such acceleration values exist at distances of $\sim$8.8 $\div$ 11 $au$.

 Figure \ref{kar24} shows a the velocity  of the ``circular'' orbits (\ref{sentr12}) vs. distance (km/s).

 The empty gaps in Figure \ref{kar24} correspond to the absence of ``circular orbits'', because at these intervals the acceleration is directed from the center.

Based on mathematical considerations, we can associate a test body at the point $M(r)$ in the ``empty band'' with a ``circular orbit'' with an imaginary center $ S' $ located on the radial line $ (SM) $ and $\overrightarrow{S'M} =-\overrightarrow{SM}$, where  $S$ is the center ( $r=0$) (see Figure \ref{kar25}).

 Figure \ref{kar25} shows a diagram explaining the bands formation mechanism corresponding to gravitation and anti-gravitation bands. Accordingly, in those bands (spherical layers) where the acceleration is directed toward the center ($A(r)<0$), a larger amount of matter will accumulate. And in the bands where ($A(r)>0$)  velocities of the test bodies $\vec{v_2}$ acquire a component that moves the body to the region from the center $S$. However, when the observational data of distant objects obtained by telescopes analyzed (stars of galactic systems) ($r\gg r_{cr}$), it is hard to estimate that the average statistical velocities in neighboring layers with different acceleration directions differ. Because of the large radii of the orbits, as well as the distances to the observer on the Earth (Figure \ref{kar25}), the angle between the vectors of these velocities $\alpha$ will be very small, and their velocity values can also be close. Therefore, for comparison with observational data, we should consider the ``observed circular rotational velocities'' (take into consideration  also  the stars in the areas of ``antigravity'', radial velocities of which are directed from the center). For the measure of these velocities, we can take the value:
 \begin{equation} \label{sentr14}
v_{co}\simeq \sqrt{r| A(r)|} ,
\end{equation}
the graph of which is shown in Figure \ref{kar26}.

In fact, in regions where ``circular orbits'' are absent (``antigravity''---region), the velocity of test bodies should be greater than in neighboring regions where they are present (``gravity''---region). This is for the reason that (in the ``antigravity'') the gravitational field accelerates the test body towards the outer region from the center. If the body was moving from the center with a velocity close to the circular to get to the right ``antigravity''---region (further from the center) its velocity should increase. To get to the left ``antigravity'' region, velocity will decrease. At the same time getting into the left ``antigravity'' region, gravitational field starts to push test body back, increasing (restoring) the velocity. Then we can assume that at the frontier $S_{2bor}$ (see Figure \ref{kar25}) (from the ``antigravity'' to the ``gravity'' region) a ``thin'' shell of test bodies with stable ``circular orbits'' is formed. You can also expect in the middle of the ``antigravity'' region the presence of a smaller number of test bodies. When the observer includes orbits of test bodies from ``antigravity'' region to the number of ``observed circular orbits'', the velocities of these bodies will be greater than in the previous ``gravity'' region.

\subsection{Numerical Solution of Equations for the Model of ``Galaxy''}
In all previous calculations, we used the mass of the Sun and the boundary conditions (\ref{sentr13}). Similar arguments and results can be obtained if we for simplicity simulate galaxies as centrally symmetric objects. To do this, we increase the mass $GM$ - $10^{7}\div 10^{11}$ times and assume that the entire mass is concentrated in the center. This point of view leads to incorrect results for small $r$ (a significant part of a galaxy mass is formed by non-central masses). However, we are interested in the behavior of the field characteristics at sufficiently large distances $r>r_{cr}$, where Kepler laws are violated. At~a~qualitative level, our reasoning does not change. Due to the multiple increase of the central mass, the critical radius $r_{cr}$ will multiply many times as well.  However, as computer calculations show, velocities of ``circular orbits'' (which in order of magnitude can be compared with what is commonly called the galaxy rotation curve) does not depend much on the mass, in contrast to the Kepler law.  Obviously, based on our paradigm, it is impossible to define the mass of a galaxy based on the peripheral bodies rotation velocity. These velocities depend on the perturbation $Z_2= Z(r_0), \ r_0$$\sim$0,  which is defined as the boundary condition. Computer experiments show that the value of $r_ {cr}$ and the amplitude of oscillatory solutions depend on the value of $Z_2$, and the oscillations period on the value of $B_n$. A~system~of units associated with kiloparsec (kpc) was used in calculations for the galaxy model.

Graphs of numerical solutions for the following parameters are given below: $k_1 =-4.035487823 \cdot 10^{-14}$  kpc$^{-2}$, $G M=G M_{\bigodot} \cdot 10^{10}  \ (kpc), \ r_0=10^{-24}$ (kpc);

\begin{equation} \label{Bn2}
B_n=0.2665 \  kpc^{-2}.
 \end{equation}

For the boundary conditions:
\begin{equation} \label{sentr15}
\alpha(r_0)=0; \  F(r_0)=0;\ Z(r_0)\equiv Z_2=1+243\cdot 10^{-8.4}; \  Z'(r_0)=0.
\end{equation}

As follows from computer studies, qualitatively, the solutions slightly depend on the boundary condition $ Z '(r_0) $. For this model, we chose this value to zero.

\begin{figure}[H]
 \centering
\includegraphics[width=0.6\textwidth]{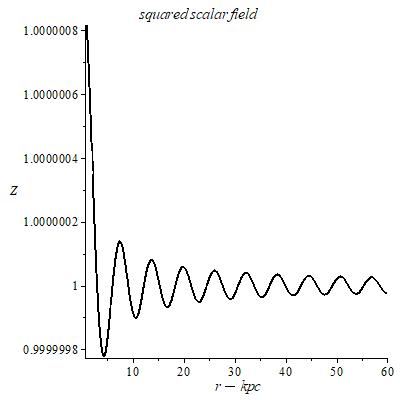}
\caption{The field Z  deviation vs. distance.
r  is measured in kiloparsec (kpc). $B_{n}=0.2665...$, $k_1 =-4.035487823 \cdot 10^{-14}, \  G M=G M_{\bigodot} \cdot 10^{10} $.}
\label{kar27}
\end{figure}

\begin{figure}[H]
 \centering
\includegraphics[width=0.6\textwidth]{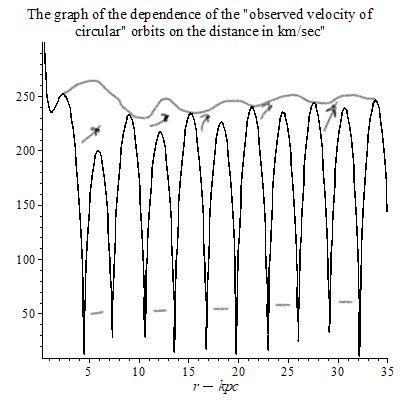}
\caption{``Observed circular rotational velocities'' vs. distance; $r \in (0.5, 25) kpc$. $B_{n}=0.2665...$, $k_1 =-4.035487823 \cdot 10^{-14}, \  GM=GM_{\bigodot} \cdot 10^{10} $.}
\label{kar28}
\end{figure}

Figures \ref{kar27} and  \ref{kar28} show the graphs of the field $Z(r)$ and ``observed circular rotational velocities''.

\begin{figure}[H]
 \centering
 \includegraphics[width=\textwidth]{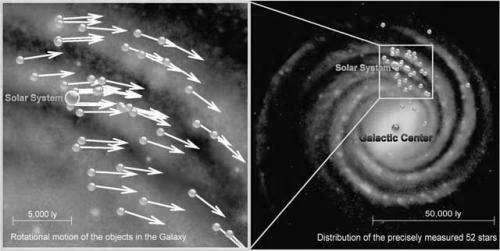}
\caption{Image of the Galaxy (Milky Way), seen from above. The distribution of 52 stars.}
\label{kar29}
\end{figure}

It is interesting to compare Figure \ref{kar25} with the Figure \ref{kar29} made on the basis of observational data~\cite{Chibueze(2014)}, as well as on the website (National Astronomical Observatory of Japan, \url{https://phys.org/news/2012-10-mass-dark-revealed-precise-milky.html}).

\begin{figure}[H]
 \centering
\includegraphics[width=0.65\textwidth]{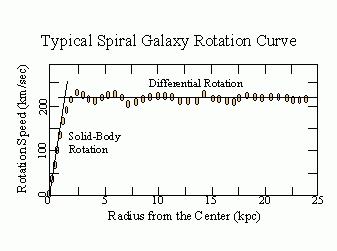}
\caption{A typical spiral galaxies rotation curve, with parameters close to the Milky Way.}
\label{kar30}
\end{figure}

In our case, for the considered parameter values, we can estimate the ``period'' $T$$\sim$6.4 kpc  (Figures~\ref{kar27} and \ref{kar28}). The order of magnitude of $T$ is similar to the order of magnitude of the periods of the rotation curves of typical spiral galaxies, shown in Figure \ref{kar30} (from the website: \url{http://astro.osu.edu/~pogge/Ast162/Unit4/spirals.html}). This graph is a simplified generalization  and does not contradict to numerous observational data cited in the works: \cite{Riess(2018),Riess(2016),Zasov(2017),Chibueze(2014)}.

 Comparison of the oscillations period $T$ of our model and the observed ``period'' $T_{gal}\approx 3.1$ kpc (Figure \ref{kar28}, and also \cite{Zasov(2017),Chibueze(2014)}  ) gives us  at least an estimate of  magnitude $B_{n}$. Proceeding from the above arguments, we propose that $T \approx 2 T_{gal}$ (taking into account the stars from $antigravity region$).  Computer experiments show that $T$ depends only on $B_n$. The problem lies in the absence of method or idea for estimating this value - the ratio of the vacuum polarization energy to $C$. The constant $C$ - has (according to our paradigm) the dimension of the squared distance. All other parameters were expressed in terms of $B_{n}$ and there is only one dimensional value in the equations  is the distance $r$. $B_{n}$ has the dimension $1/r^2$ (\ref{min4}).

\subsection{General Characteristics of the Models. Attempt to Reconcile}

In all three models (the cosmological model, the conditional ``Sun'' model, the galaxy model) various systems of units were used: centimeter or Hubble unit ($hu$) equal to $t_m$$\sim$13.8 $\cdot 10^{9}$ years, astronomical unit $au$, kiloparsec $kpc \approx 2.062706271\cdot 10^{8}$ $au$. The values of $ B_ {n} $ for these three different scales was chosen, on the one hand, on the basis of the principle of apparent consistency with observational data, and on the other hand, on the basis of computer capabilities. The question arises: Is there an unaccounted mechanism that leads to different values of $B_{n}$ for theories corresponding to different scales and different objects? We leave this question for further research.

 Let's consider what happens when we set $B_{n0} = 0.2665...$ kpc$^{-2} \approx 4.770463302 \cdot 10^{12}$ hu$^{-2} $  as the basis, which corresponds to the model of the galaxy and which leads to oscillations with period \scalebox{0.95}[0.95]{$T \approx$ 6.2 kpc}  (Figure \ref{kar26}). Let us compare this period with the period of oscillations (the scale factor) in the cosmological model. By solving equations for the cosmological model, we can numerically find the period of oscillations for the value $B_{n0}$ for  $x=1\cdot hu$:  $ \ T_{h1}\approx 17,000$ Earth years.
 Note that for the time $ \ T_{h1}\approx 17,000$ the light travels the distance $\sim$5.31 kpc. Thus, the scale of the galactic periodic structures and the scale of the oscillations of the entire universe (considered as a homogeneous and isotropic space-time) are similar, and possibly match. Perhaps this fact is a clue to the development of a theory describing the universe and spatial structures from the unified point of view.

\begin{figure}[H]
 \centering
 \includegraphics[width=7cm]{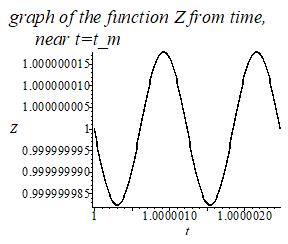}
\caption{$Z(t)$---solutions of Equations (\ref{furrav6}) and (\ref{furrav7}) for time interval that equals to two periods of oscillations $\bigtriangleup t = 0.0000024750$ hu.}
\label{kar31}
\end{figure}

Figure \ref{kar31} shows the graph of the function $Z=Z(t)$---solving the Equations (\ref{furrav6}) and (\ref{furrav7}) for a time equal to two periods of oscillation $\bigtriangleup t=0.0000024750$ hu.

 The result is a bit unexpected. Earlier (see Figures \ref{kar1}, \ref{kar16} and \ref{kar17}), (and in the work \cite{Zaripov(2014)}), we~predicted $T \sim 10^{-1} - 10^{-2}$ hu. Also, the authors of  \cite{Ringermacher(2015)} reported the discovery of cosmological fluctuations (from the analysis of observational data) of the scale factor, with a period of  $T$$\sim$  1 hu/7. For the description of dark matter, a scalar field model used in this paper as well.  However, the question arises whether an analysis of observational data takes into account the fact that the ``Hubble~constant''  fluctuates. Light travels from galaxies millions and billions of years. We observe it in the past and during that time there were many periods of oscillation. Perhaps, we need to revise the linear Hubble law (redshift), to take into account fluctuations comparable to the mean value. It is possible that the oscillations contain other harmonics of fluctuations corresponding to different scales. As pointed out by J. Lemaitre, the linear correlation between the cosmological redshifts (caused by the expansion of the universe) and distances is by no means absolute.

 In the cosmography method proposed in Ref. \cite{Aviles(2012)}, observational cosmological data are computed by the redshift parameter z. In our case, when the solutions are oscillatory in nature, the inverse dependence of the Hubble constant on cosmological time is not unambiguous within the limits of observations accuracy (also calculations related by approximation in the model). Such a one-to-one correspondence can be traced between the scale factor and cosmological time in the later stages of the Universe expansion. To interpret the observational data with regard to these effects, additional understanding of the theoretical assumptions and methods of obtaining observational data on the ``red shift'' is required.

When converting the value of $B_{n0}$ into astronomical units, we get a small value $B_{n0}=0.2665...kpc^{-2}\approx  6.263577 \cdot 10^{-18} au^{-2}$ instead of $B_{n}\approx 0.0059986 au^{-2}$  considered in the  conditional "Sun" model.  The oscillation period depends on $B_{n}$ and is independent of the mass (the harmonic spectrum depends on mass). Thus, if the parameter $B_{n}$ is one for the entire universe and is similar to $B_{n0}$, then in the immediate  vicinity of the solar system there is no gravitational field with ``antigravity'' regions caused by the Sun.

\begin{figure}[H]
 \centering
\includegraphics[width=8cm]{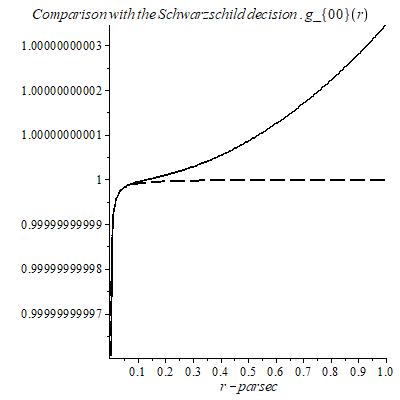}
\caption{A comparison of the numerical solution $g_{00}$ (solid line) with the Schwarzschild-de Sitter solution (dashed
line); r---in the parsecs. $B_{n}=1.599 \cdot 10^{-6}  (pc^{-2})$. The solid graph is obtained for the boundary condition: $Z(0)$ = 1.00009833247, $p(0)$ = 0, $\alpha(0)$ = 0, $F(0)$ = 0.
}
\label{kar32}
\end{figure}

\begin{figure}[H]
 \centering
 \includegraphics[width=8cm]{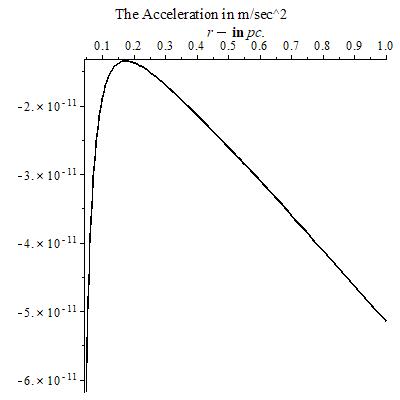}
\caption{Test body acceleration $ A (r) $  vs. distance; r---in the parsecs. For the case of solar mass. $B_{n}=1.599 \cdot 10^{-6}  (pc^{-2})$, $Z(0)$ = 1.00009833247, $p(0)$ = 0, $\alpha(0)$ = 0, $F(0)$ = 0.
}
\label{kar33}
\end{figure}

Let's return to the model of the conditional ``Sun'' at $B_{ n}=B_{n0}$. Differences from the Newtonian law of gravity appear at  $r_1 > 0.1 - 0.5 pc$ (unit of measure $r$ parsec). At such distances, there is an additional acceleration directed toward the center, with the value of about  $10^{-11}m/c^2$. With the increasing $r$, the acceleration of the test body (directed toward the center) increases monotonically for $r>r_1$. The amplitude of the deviation (value of the acceleration) depends on the value of the perturbation $Z_2$. Here is the comparative graph of the metric component  $g_{00}$ for our case and for the Schwarzschild metric ($Z=const$) (Figure \ref{kar32}), as well as the acceleration graph $ A (r) $ (Figure \ref{kar33}). The ``reversal'' point of the graph $ A(r)$ corresponds to the point $ \nu''(r_ {cr}) = 0 $.

Apparently, we can talk about the synchronized gravitational field that connects all objects of the galaxy's system.

The team of authors  (van Dokkum   et al. ) recently published velocity measurements of luminous globular clusters in the galaxy NGC 1052-DF2, concluding that it lies far off the canonical stellar mass---halo mass (SMHM) relation \cite{Dokkum(2017),Dokkum(2018),P_Dokkum(2018)}. They report that they have discovered a galaxy in which dark matter is almost completely absent. On the other hand, there are a lot of data on other galaxies, indicating that they almost entirely consist of dark matter \cite{Dokkum(2016),Riess(2016),Planck(2016)}. Computer experiments allow us to hope that such a difference in the observational data can be modeled within the framework of the MTIG by varying the parameters $Z_2, GM$.

%%%%%%%%%%%%%%%%%%%%%%%%%%%%%%%%%%%%%%%%%%
\section{Conclusions}

The main results of our research.
\begin{quote}

MTIG model is proposed for a macroscopic description of gravity and cosmology, which (possibly) is capable of solving problems 1--3, given at the beginning of the article, and motivating to further experiments. We propose the working hypothesis according to which the physical parameters associated with gravitation, such as the gravitational and cosmological ``constants'' - $G$ and $\Lambda_{eff}$, the Hubble ``constant'' $H$, in addition to monotonic evolution, fluctuate about their mean values. Because of the implementation of the two branches of solutions, these fluctuations can contain elements of stochasticity. This hypothesis is realized in the mathematical model considered in this article.
\end{quote}
\begin{quote}
The solutions of the MTIG equations for the case of a centrally symmetric gravitational field, in addition to the Schwarzschild-de Sitter solutions (for $Z=const$), contain solutions that lead to new physical effects at large distances from the center.   For distances greater than a certain critical value, the following effects can appear (depending on the value of the parameters of the theory): deviation from the law of gravitational interaction of general relativity and its Newtonian approximation, antigravity, absence of asymptotic flatness at infinity. The responsibility for these effects (in the first place) carries not an integration constant that corresponds to the mass of a physical body, but some other (dimensionless) charge $Z_2 $, which is brought into the theory as the boundary value of the field $Z(r)$ at the center of symmetry ($r=0$), different from  the (vacuum) mean value ($Z_0=1$) of  field $Z$. The mass affects the value of the critical radius, the charge $Z_2$ affects the amplitude of the deviation, $B_n$ affects the oscillation frequency. We consider the hypothesis that the parameter $B_n$ is the same for all objects, and the charge $Z_2$ is different for local objects (stars, galaxies, clusters...).

 Thus, the flat asymptotics (the asymptotics of the field at large distances  $r > r_{cr}$ in the Newton and Schwarzschild theory) becomes unstable and can enter the oscillatory regime with the elements of of chaotic behavior. The mean oscillation frequency depends on the value of $ B_n $, which depends on the vacuum polarization energy density. Near the center, the influence of mass predominates and the laws of general relativity and Newton's law of universal gravitation (for the weak field) are approximately fulfilled. At sufficiently large distances (greater than critical) these solutions pass into other solutions, where the influence of $ Z_2 $ and $ B_n $ prevails.   (Because of the nonlinearity of the theory,  such allocation of roles of the parameters, based on computer experiments, is conditional (approximate)).

\end{quote}
%%%%%%%%%%%%%%%%%%%%%%%%%%%%%%%%%%%%%%%%%%
\section{Discussion}
In the general case, the following working hypothesis is advanced.

The consideration of the solutions of the MTIG equations in homogeneously isotropic and in centrally symmetric space-time leads to the model claiming a single description for DM and DE. It is assumed that the scalar field $Z=Z(r,t)$ plays the role of a certain World framework (glue) that synchronizes cosmological evolution with the existing mass type local objects.
The observed effects of dark energy and dark matter are manifestations of proposed by us mechanism.

Based on the data that dark matter is observed on galactic scales (large-scale inhomogeneities), we estimated the parameter $B_n$. Fluctuations of the scale factor of the cosmological model also have  been agreed with this value. If we assume that this parameter is unified, then for the solar system the deviations associated with the phenomenon of dark matter should manifest themselves at distances $r_1 > 0.01 - 0.5$ pc (depending on the value of $Z_2$ attributed to the Sun) as an additional acceleration to the Sun. The objects of the Oort Cloud of the Solar System correspond to such distances. It is possible that changes in the parameters of long-period comets are related to the phenomenon under~consideration.

The notion of dark matter - that it consists of WIMP's, massive particles that almost do not interact with particles of ordinary matter, in our opinion, probably does not oppose the proposed theory of MTIG. MTIG is a classical, phenomenological description, based on a nonlinear theory and in the future it is necessary to relate it to a quantum description. The connection between MTIG and the scale invariance of quantum field theories and the violation of this invariance seemed to us a promising direction. We noticed that in MTIG, instabilities leading to chaotic solutions are closely related to the scale invariance of the equations.

We have not finished the study of solving equations in a centrally symmetric space when the field $ Z $ is not static: $ Z = Z (r, t) $. These studies will be continued.

%%%%%%%%%%%%%%%%%%%%%%%%%%%%%%%%%%%%%%%%%%
\vspace{5pt}
%%%%%%%%%%%%%%%%%%%%%%%%%%%%%%%%%%%%%%%%%%
%%%%%%%%%%%%%%%%%%%%%%%%%%%%%%%%%%%%%%%%%%
\funding{``This research received no external funding'' }
% please add this part.

\abbreviations{The following abbreviations are used in this manuscript:\\

\noindent
\begin{tabular}{@{}ll}
MDPI & Multidisciplinary Digital Publishing Institute\\
DOAJ & Directory of open access journals\\
TLA & Three letter acronym\\
LD & linear dichroism\\
MTIG & Modified Theory of Induced Gravity
\end{tabular}}

%=====================================
% References, variant A: internal bibliography
%=====================================
\reftitle{References}

% The following MDPI journals use author-date citation: Arts, Econometrics, Economies, Genealogy, Humanities, IJFS, JRFM, Laws, Religions, Risks, Social Sciences. For those journals, please follow the formatting guidelines on http://www.mdpi.com/authors/references
% To cite two works by the same author: \citeauthor{ref-journal-1a} (\citeyear{ref-journal-1a}, \citeyear{ref-journal-1b}). This produces: Whittaker (1967, 1975)
% To cite two works by the same author with specific pages: \citeauthor{ref-journal-3a} (\citeyear{ref-journal-3a}, p. 328; \citeyear{ref-journal-3b}, p.475). This produces: Wong (1999, p. 328; 2000, p. 475)

%=====================================
% References, variant B: external bibliography
%=====================================
%\externalbibliography{yes}
%\bibliography{your_external_BibTeX_file}

\begin{thebibliography}{999}

\bibitem{Weinberg C.(1989)}
Weinberg, C.S. The cosmological constant problem. \emph{Rev. Mod. Phys.} \textbf{1989}, \emph{61}, 1--23.

\bibitem{Speake(2014)}
Speake, C.;  Quinn, T.  The search for Newton’s constant.  \emph{Phys. Today} \textbf{2014}, \emph{67},  27--33.

\bibitem{Rosi(2014)}
Rosi, G.; Sorrentino, F.; Cacciapuoti, L.; Prevedelli, M.; Tino G.M. Precision measurement of the Newtonian gravitational constant using cold atoms.  \emph{Nature} \textbf{2014}, \emph{510}, 518--521.


\bibitem{Luo(2018)}
Luo, J.; Qing, L.; Chao, X.; Qi, L.; Hao, X.; Hu, Z.-K.; Wu, S.-C.; Milyukov, V. Measurements of the gravitational constant using two independent method. \emph{Nature} \textbf{2018}, \emph{560}, 582--588, doi:10.1038/s41586-018-0431-5.

\bibitem{Riess(2018)}
Riess, A.G.; et al.
New Parallaxes of Galactic Cepheids from Spatially Scanning the Hubble Space Telescope: Implications for the Hubble Constant. \emph{arXiv} \textbf{2018}, arXiv:1801.01120.
% Please include the first ten authors' names before using et al. in references. and the followings are same.

\bibitem{Riess(2016)}
Riess, A.G.; Macri, L.M.; Hoffmann, S.L.; et al. Determination of the Local Value of the Hubble Constant.  \emph{Astrophys. J.} \textbf{2016}, doi:10.3847/0004-637X/826/1/56.

\bibitem{Planck(2016)}
Planck, C.; Aghanim, N.; Ashdown, M.; et al. Planck intermediate results XLVI. Reduction of large-scale systematic effects in HFI polarization maps and estimation of the reionization optical depth. \emph{Astron. Astrophys.}  \textbf{2016}, \emph{596}, A107.

\bibitem{Rham(2007)}
de Rham,  C.; Hofmann, S.; Khoury, J.; Tolley, A.J. Cascading Gravity and Degravitation. \emph{arXiv} \textbf{2007}, arXiv:0712.2821.


\bibitem{Dvali(2000)}
 Dvali, G.R.; Gabadadze, G.; Porrati, M.  4D Gravity on a Brane in 5D Minkowski Space. \emph{Phys. Lett. B} \textbf{2000}, \emph{485}, 208--214.

\bibitem{Dvali(2001)}
 Dvali, G.R.; Gabadadze, G. Gravity on a Brane in Infinite-Volume Extra Space. \emph{Phys. Rev. D} \textbf{2001}, \emph{63}, 065007--065020.

 \bibitem{Dvali(2007)}
 Dvali, G.R.; Hofmann, S.;   Khoury, J. Degravitation of the Cosmological Constant and Graviton Width. \emph{Phys. Rev. D} \textbf{2007}, \emph{76}, 084006.

 \bibitem{Ravanpak(2016)}
 Ravanpak, A.; Farajollahi, H.; Fadakar,  G.F. Probing Lambda-DGP Braneworld Model. \emph{Astron. Astrophys.} \textbf{2016}, \emph{16}, 137.

\bibitem{Capozziello(2018)}
Capozziello, S.; Luongo, O.; Pincak, R.;  Ravanpak, A. Cosmic acceleration in non-flat f(T) cosmology. \emph{arXiv} \textbf{2018}, arXiv:1804.03649.

\bibitem{Arkani(2002)}
 Arkani-Hamed, N.; Dimopoulos, S.;  Dvali, G.;  Gabadadze, G. Non-Local Modification of Gravity and the Cosmological Constant Problem. \emph{arXiv} \textbf{2002}, arXiv:0209227.

\bibitem{Mishra(2017)}
 Mishra, S.S.; Sahni, V.; Shtanov, Y. Sourcing Dark Matter and Dark Energy from $\alpha$-attractors. \emph{arXiv} \textbf{2017}, arXiv:1703.03295.

\bibitem{Sahni(2000)}
Sahni, V.;  Starobinsky, A.A. The case for a positive cosmological  $\Lambda$-term. \emph{Int. J. Mod. Phys. D} \textbf{2000}, \emph{9}, 373--443.

\bibitem{Peebles(2003)}
Peebles, P.J.E.; Ratra, B. The cosmological constant and dark energy. \emph{Rev. Mod. Phys.} \textbf{2003}, \emph{75}, 559--606.

\bibitem{Padmanabhan(2003)}
Padmanabhan, T. Cosmological constant---The weight of the vacuum. \emph{Phys. Rep.} \textbf{2003}, \emph{380}, 235--320.

\bibitem{Sahni(2006)}
Sahni, V.;  Starobinsky, A.A. Reconstructing Dark Energy. \emph{Phys. D} \textbf{2006}, \emph{15}, 2105--2132.

\bibitem{Copeland(2006)}
Copeland, E.J.; Sami, M.;  Tsujikawa, S. Dynamics of dark energy. \emph{Int. J. Mod. Phys. D} \textbf{2006}, \emph{15}, 1753--1936.

\bibitem{Bousso(2008)}
Bousso, R. The cosmological constant. \emph{Gen. Relativ. Gravit.} \textbf{2008}, \emph{40}, 607--637.

\bibitem{Kallosh(2013)}
Kallosh, R.; Linde, A.  Universality Class in Conformal Inflation. \emph{J. Cosmol. Astropart. Phys.} \textbf{2013}, \emph{1307},
002.

\bibitem{Kallosh(2015)}
Kallosh, R.; Linde, A.  Multi-field Conformal Cosmological Attractors. \emph{arXiv} \textbf{2015}, arXiv:1309.2015.



\bibitem{Zaripov(2007)}
Zaripov, F.S. A conformally invariant generalization of string theory to higher-dimensional objects. Hierarchy of coupling constants. \emph{Gravit. Cosmol.} \textbf{2007}, \emph{13},  273--281.

\bibitem{Zaripov(2014)}
Zaripov, F. Modified equations in the theory of induced gravity. \emph{Astr. Space Sci.} \textbf{2014}, \emph{352},  289--305.

\bibitem{Zaripov(2017)}
Zaripov, F.S.  Phenomenological Model of Multiphase Cosmological Scenario in Theory of Induced Gravity. \emph{Russ. Phys. J.} \textbf{2017}, \emph{59}, 1834--1841.


\bibitem{De Felice(2010)}
De Felice, A.;  Tsujikawa, S. f(R) theories. \emph{Living Rev. Relativ.} \textbf{2010}, doi:10.12942/lrr-2010-3.

\bibitem{Nojiri(2007)}
 Nojiri, S.I.; Odintsov, S.D. Introduction to Modified Gravity and Gravitational Alternative for Dark Energy, \emph{Int. J. Geom. Methods Mod. Phys.} \textbf{2007}, \emph{4}, 115--145.

 \bibitem{Nojiri(2014)}
 Nojiri, S.I.; Odintsov, S.D. Accelerating cosmology in modified gravity: From convenient F(R) or string-inspired theory to bimetric F(R) gravityInt. \emph{J. Geom. Methods Mod. Phys.} \textbf{2014}, \emph{11},  1--24.

\bibitem{Peter(2016)}
Peter, K.; Dunsby, S.; Luongo, O.; Reverberi, L. Dark Energy and Dark Matter from an additional adiabatic fluid. \emph{Phys. Rev. D} \textbf{2016}, \emph{94}, 083525.

\bibitem{Luongo(2018)}
Luongo, O.; Muccino, M. Speeding up the Universe using dust with pressure. \emph{Phys. Rev. D}  \textbf{2018}, \emph{98}, 103520.

\bibitem{Aviles(2012)}
 Aviles, A.; Gruber, C.; Luongo,O.; Quevedo, H.Cosmography and constraints on the equation of state of the Universe in various parametrizations. \emph{Phys. Rev. D} \textbf{2012}, \emph{86}, 123516.

 \bibitem{Sakharov(1968)}
Sakharov, A.D. Vacuum Quantum Fluctuations In Curved Space And The Theory Of Gravitation.  \emph{Sov. Phys. Dokl}.\textbf{1968}, \emph{12}, 1040.

\bibitem{Visser(2002)}
Visser, M. Sakharov’s Induced Gravity: A Modern Perspective. \emph{Mod. Phys. Lett. A} \textbf{2002}, \emph{17}, 977.

\bibitem{Andrianov(2006)}
Andrianov, A.A.; Andrianov, V.A.; Giacconi, P.;  Soldati,  R.  Induced gravity and universe creation on the domain wall in five-dimensional space-time. \emph{Theor. Math. Phys.} \textbf{2006}, \emph{148}, 880.

\bibitem {Linnemann(2018)}
Linnemann, N.S.;  Visser,  M.R. Hints towards the Emergent Nature of Gravity. \emph{arXiv} \textbf{2018}, arXiv:1711.10503v2.

\bibitem {Scholz(2011)}
Scholz, E. Weyl  geometry in late 20th century physics. emph{arXiv} \textbf{2011}, arxiv:1111.3220v1.

\bibitem {Aalbers(2013)}
Aalbers, J. Conformal Symmetry in Classical Gravity. 2013. Available online: \url{http://dspace.library.uu.nl/handle/1874/280136} 

\bibitem {Dengiz(2011)}
Dengiz, S.; Tekin, B. Higgs Mechanism for New Massive Gravity and Weyl Invariant Extensions of Higher Derivative Theories.  \emph{Phys. Rev. D} \textbf{2011}, \emph{84}, 024033.

\bibitem {Carballo-Rubio(2015)}
Carballo-Rubio, R. Longitudinal diffeomorphisms obstruct the protection of vacuum energy. \emph{Phys. Rev. D}   \textbf{2015}, \emph{91}, 124071.

\bibitem{Kamenshchik(2016)}
Kamenshchik, AY.;  Pozdeeva,  E.O.; Starobinsky, A.A.;  Tronconi, A.; Venturi, G.;  Vernov, S.Y.  Transformations between Jordan and Einstein frames: Bounces, antigravity, and crossing singularities. \emph{Phys. Rev. D}  \textbf{2016}, \emph{94}, 063510.

\bibitem{Bars(2016)}
Bars, I.; James A.  Physical interpretation of antigravity. \emph{Phys. Rev. D}  \textbf{2016},  \emph{93}, 044029.

 \bibitem{Grin(1988)}
 Green, M.; Schwarz, J.;  Witten, E. \emph{Superstring Theory}; Cambridge University Press:  Cambridge, UK, 1988; p. 518.
% newly added information, please confirm.

 \bibitem{Regge(1977)}
 Regge, T.; Teitelboim, C. General Relativity à la string: A progress report. In Proceedings of the First Marcel Grossmann Meeting, Trieste, Italy, 1975.
% newly added information, please confirm.

 \bibitem{Paston(2010)}
 Paston, S.A.; Semenova, A.N. Constraint algebra for Regge-Teitelboim formulation of gravity. \emph{Int. J. Theor. Phys.} \textbf{2010}, \emph{49}, 2648--2658.

 \bibitem{Sheykin(2014)}
 Sheykin, A.A.; Paston, S.A. The approach to gravity as a theory of embedded surface. \emph{AIP Conf. Proc.} \textbf{2014}, \emph{1606}, 400.

 \bibitem{Stephani(2003)}
 Stephani, H.; Kramer, D.; MacCallum, M.; Hoenselaers, C.; Herlt, E. \emph{Exact Solutions of Einstein’s Field Equations}, 2nd ed.; Cambridge Monographs on Mathematical Physics; Cambridge University Press: Cambridge, UK, 2003; p. 25.

\bibitem{Bamba(2012)}
 Bamba, K.; Capozziello, S.;  Nojiri, S.; Odintsov, S.D. Dark energy cosmology: the equivalent description via different theoretical models and cosmography tests. \emph{Astrophys. Space Sci.} \textbf{2012}, \emph{342}, 155--228.

\bibitem{Rham(2008)}
 de Rham, C.; Dvali, G.; Hofmann, S.; Khoury,J.; Pujolàs,O.; Redi, M.;  Tolley, A.J. Cascading Gravity: Extending the Dvali-Gabadadze-Porrati Model to Higher Dimension. \emph{Phys. Rev. Lett.} \textbf{2008}, \emph{100}, 251603.

\bibitem{Zaripov(1995)}
Zaripov, F.SH.  A conformally invariant sigma-model as generalization of the theory of strings. In Proccedings of the International School-Seminar  of a Foundation of the Theory of a Gravitation and Cosmology, Odessa, Ukraine, 1995; p. 35.
% newly added information, please confirm.


\bibitem{Zaripov(1986)}
Zaripov, F.S. On the stability of the Friedman world with a charged scalar field and with self-action. In \emph{Gravitation and Theory of Relativity}; KSU, Publishers: Kazan, Russia, 1986; pp. 62--74. (In Russian)

\bibitem{Zaripov(2010)}
Zaripov, F.S. Generalized equations of induced gravity. The evolution of coupling constants.
 \emph{Vestnik TGGPU} \textbf{2010},  \emph{4}, 23--28. (In Russian)

\bibitem{Chervon(1997)}
 Chervon, S.V.; Zhuravlev, V.M.; Shchigolev, V.K. New exact solutions in standard inflationary models. \emph{Phys. Lett. B} \textbf{1997}, \emph{398},  269--273.

 \bibitem{Zhuravlev(1998)}
 Zhuravlev, V.M.;  Chervon, S.V.;  Shchigolev, V.K. \emph{New Classes of Exact Solutions in Inflationary Cosmology}; JETF: New York, NY, USA, 1998; Volume 87, pp. 223--228.

\bibitem{Buchbinder(1992)}
Buchbinder, I.L.;  Odintsov, S.D.; Shapiro, I.L. \emph{Efective Action in Quantum Gravity}; IOP: Bristol, PA, USA, 1992; p. 413.

\bibitem{Gorbunov(2014)}
Gorbunov, D.;  Tokareva, A. Scale-invariance as the origin of dark radiation? \emph{Phys. Lett. B} \textbf{2014}, \emph{739}, 50--55.

\bibitem{Rham(2014)}
de Rham, C. Massive Gravity. \emph{Living Rev. Relativ.} \textbf{2014}, \emph{17}, 1--189.

\bibitem{Clifton(2012)}
 Clifton, T.; Ferreira, P.G.; Padilla, A.; Skordis, C. Modified Gravity and Cosmology. \emph{arXiv} \textbf{2012}, arXiv:1106.2476v3.


\bibitem{Lerner(2018)}
Lerner, E.J. Observations contradict galaxy size and surface brightness predictions that are based on the expanding universe hypothesis. \emph{arXiv} \textbf{2018}, arXiv:1803.08382.


\bibitem{Coleman(1977)}
 Coleman, S. Fate of the false vacuum: Semiclassical theory. \emph{Phys. Rev. D}  \textbf{1977}, \emph{15}, 2929--2936.

\bibitem{Coleman(1980)}
Coleman, S.; De Luccia, F. Gravitational effects on and of vacuum decay. \emph{Phys. Rev. D} \textbf{1980}, \emph{21}, 3305--3315.

\bibitem{Lee(1987)}
  Lee, K.; Weinberg, E.J. Decay of the true vacuum in curved space-time. \emph{Phys. Rev. D}  \textbf{1987}, \emph{36}, 1088--1094.

\bibitem{Hackworth(2004)}
Hackworth, J.C.;  Weinberg,  E.J. Oscillating bounce solutions and vacuum tunneling in de Sitter spacetime. \emph{Phys. Rev. D}  \textbf{2004}, \emph{71}, 044014--044032.

 \bibitem{Masoumi(2016)}
Masoumi, A.; Paban, S.; Weinberg, E.J. Tunneling from a Minkowski vacuum to an AdS vacuum: A new thin-wall regime. \emph{Phys. Rev. D} \textbf{2016}, \emph{94}, 025023--025040.

\bibitem{Linde(2003)}
Linde, A. \emph{Inflation, Quantum Cosmology and the Anthropic Principle};
Cambridge University Press: Cambridge, UK, 2003; pp. 1--35.
% newly added information, please confirm.

\bibitem{Anderson(2002)}
Anderson, J.D.;  Laing, P.A.; Lau, E.L.;  Nieto,  M.M.; Turyshev, S.G. Search
for a Standard Explanation of the Pioneer Anomaly. \emph{Mod. Phys. Lett. A} \textbf{2002}, \emph{17},
875--885.

\bibitem{Nieto(2005)}
Nieto, M.M.; Anderson, J.D. Using Early Data to Illuminate the Pioneer Anomaly. \emph{Class. Quant. Grav.} \textbf{2005}, \emph{22},  5343--5354.

\bibitem{Chibueze(2014)}
Chibueze, J.O.;  Sakanoue, H.; Nagayama, T.; Omodaka, T.; Handa, T.; et al. Trigonometric parallax of IRAS 22,555 + 6213 with VERA: Three-dimensional view of sources along the same line of sight. \emph{Astron. Soc. Jpn.} \textbf{2014}, \emph{66}, 104.

\bibitem{Zasov(2017)}
Zasov, A.V.; Saburova, A.S.; Khoperskov, A.V.; Khoperskov, S.A. Dark matter in galaxies.  \emph{Phys. Uspekhi} \textbf{2017}, \emph{60}, 3--39..

\bibitem{Ringermacher(2015)}
Ringermacher, H.I.; Mead, L.R. Observation of Discrete Oscillations in a Model-Independent Plot of Cosmological Scale Factor versus Lookback Time and Scalar Field Model. \emph{Astron. J.} \textbf{2015}, \emph{149},  137.

\bibitem{Dokkum(2017)}
 van Dokkum, P.; Abraham, R.; Romanowsky, A.J.; Brodie, J.; Conroy, C.; Danieli, S.; Zhang, J. Extensive globular cluster systems associated with ultra diffuse galaxies in the Coma cluster. \emph{Astrophys. J. Lett.} \textbf{2017}, \emph{844}, L11.

 \bibitem{Dokkum(2018)}
van Dokkum, P.; et al. A galaxy lacking dark matter. \emph{Nature} \textbf{2018}, \emph{555}, 629--632.

\bibitem{P_Dokkum(2018)}
van Dokkum,  P.; Cohen, Y.; Danieli, S.; et al. An Enigmatic Population of Luminous Globular Clusters in a Galaxy Lacking Dark Matter. \emph{Astrophys. J. Lett.} 2018, \emph{856}, L30.

\bibitem{Dokkum(2016)}
van Dokkum, P; Abraham, R.; Brodie,J; Conroy, C.; Danieli, S.; et al. A High Stellar Velocity Dispersion and ~100 Globular Clusters for the Ultra Diffuse Galaxy Dragonfly 44. \emph{Astrophys. J. Lett.} \textbf{2016}, \emph{828}, L6.







\end{thebibliography}

%%%%%%%%%%%%%%%%%%%%%%%%%%%%%%%%%%%%%%%%%%
%% optional
%\sampleavailability{\hl{Samples} of the compounds ...... are available from the authors.}
% please confirm if this part is necessary, if not, please delete.

%% for journal Sci
%\reviewreports{\\
%Reviewer 1 comments and authors’ response\\
%Reviewer 2 comments and authors’ response\\
%Reviewer 3 comments and authors’ response
%}

%%%%%%%%%%%%%%%%%%%%%%%%%%%%%%%%%%%%%%%%%%
\end{document}